\documentclass[prx,twocolumn,showpacs,superscriptaddress,floatfix]{revtex4-2}
\usepackage{amsmath}
\usepackage{bm}
\usepackage{graphicx}
\usepackage[ansinew]{inputenc}
\usepackage{array}
\usepackage{color}
\usepackage{amsxtra} 
\usepackage{amstext}
\usepackage{amssymb}
\usepackage{latexsym}
\usepackage{dsfont}
\usepackage{braket}
\usepackage[caption=false]{subfig}
\usepackage[makeroom]{cancel}
\usepackage{lineno}

\usepackage{dcolumn}
\usepackage{bm}
\usepackage{bbm}
\usepackage{braket}
\usepackage{mathrsfs}
\usepackage{amstext}
\usepackage{euscript}
\usepackage{txfonts}

\newcommand{\ms}[1]{\mbox{\scriptsize #1}}
\newcommand{\msi}[1]{\mbox{\scriptsize \textit{#1}}}

\begin{document}

\title{Quantum theory of single-photon nonlinearities generated by ensembles of emitters}

\author{Kurt Jacobs}
\affiliation{United States Army Research Laboratory, Adelphi, Maryland 20783, USA}
\affiliation{Department of Physics, University of Massachusetts at Boston, Boston, Massachusetts 02125, USA}

\author{Stefan Krastanov} 
\affiliation{Department of Electrical Engineering and Computer Science, Massachusetts Institute of Technology, Cambridge, MA 02139, USA}

\author{Mikkel Heuck}
\affiliation{Department of Electrical and Photonics Engineering, Technical University of Denmark, 2800 Lyngby, Denmark}

\author{Dirk R. Englund}% 
\affiliation{Department of Electrical Engineering and Computer Science, Massachusetts Institute of Technology, Cambridge, MA 02139, USA}

\date{\today}

\begin{abstract} 
The achievement of sufficiently fast interactions between two optical fields at the few-photon level would provide a key enabler for a broad range of quantum technologies. One critical hurdle in this endeavor is the lack of a comprehensive quantum theory of the generation of nonlinearities by ensembles of emitters. Distinct approaches applicable to different regimes have yielded important insights: i) a semiclassical approach reveals that, for many-photon coherent fields, the contributions of independent emitters add independently allowing ensembles to produce strong optical nonlinearities via EIT; ii) a quantum analysis has shown that in the few-photon regime collective coupling effects prevent ensembles from inducing these strong nonlinearities. Rather surprisingly, experimental results with around twenty photons are in line with the semi-classical predictions. Theoretical analysis has been fragmented due to the difficulty of treating  nonlinear many-body quantum systems. Here we are able to solve this problem by constructing a powerful theory of the generation of optical nonlinearities by single emitters and ensembles. The key to this construction is the application of perturbation theory to perturbations generated by subsystems. This theory reveals critical properties of ensembles that have long been obscure. The most remarkable of these is the discovery that quantum effects prevent ensembles generating single-photon nonlinearities only within the rotating-wave regime; outside this regime single-photon nonlinearities scale as the number of emitters. The theory we present here also provides an efficient way to calculate nonlinearities for arbitrary multi-level driving schemes, and we expect that it will prove a powerful foundation for further advances in this area.
\end{abstract}

\maketitle

%\linenumbers

\section{Introduction} 

The realization of optical nonlinearities strong enough to perform logic operations at the single photon level is a long-standing goal in quantum information science and technology. Such ``giant'' nonlinearities would have a range of applications including fast all-optical classical and quantum information processing~\cite{Langford11, Brod16, Niu18, Niu18b, Heuck20, Li20, Krastanov21} and the production of complex non-classical states for quantum sensing~\cite{Wang21, Johnsson20, Zhou18, Zhang14}. 

Atoms and other emitters will induce nonlinearities for electromagnetic fields when the frequencies of the fields are sufficiently off-resonant with the atomic transitions that the effect of the light on the emitters is perturbative~\cite{Bloembergen64, Oudar80, Boyd81, Szymanowski94, Harris90, FIM05}. Under this condition, the atomic polarization is given by a power series in the field amplitudes. Each term in this power series is an induced nonlinearity of a given order. There is a fundamental limitation to the strength of these nonlinearities. Since the effect of the light on the emitter must be perturbative the emitter/field coupling, $g$, must be small compared to the detuning, $\Delta$, between the light sources and the emitter transitions to which they couple. As we will see below, terms in the perturbation expansion corresponding to an $n^{\msi{th}}$-order nonlinearity are proportional to $g \epsilon^{n}$ with $ \epsilon \equiv g/\Delta \ll 1$. All nonlinearities generated by a single atom are therefore much smaller than the emitter/field coupling rate, $g$, and decrease exponentially with $n$, the order of the nonlinearity. For an \textit{ensemble} of $N$ independent emitters, a semiclassical analysis, valid for sufficiently strong coherent fields, shows that the non-linearities induced by each emitter simply sum together to produce $N$ times the nonlinearity of a single emitter. This property allows ensembles to circumvent the above bound and generate ``giant" nonlinearities~\cite{Bloembergen64, Oudar80, Boyd81, Szymanowski94, Harris90, FIM05}. By employing the technique of electromagnetically induced transparency (EIT) the nonlinearities can be generated without inducing significant dissipation for the fields~\cite{Harris90, FIM05}. 

In 1997 Imamo{\=g}lu \textit{et al.\ }asked whether an ensemble of $N$ atoms could similarly be used to generate giant nonlinearities for a \textit{single photon} in a cavity~\cite{Imamoglu97}. Due to the difficulty of treating the quantum dynamics of a dissipative ensemble of emitters coupled to a cavity mode all analyses performed at the time involved reasoning from a simplified version of the system. It was nevertheless concluded that the answer was negative, due at least partially to the sharpness of the EIT linewidth as compared to that of an optical cavity~\cite{Grangier98, Gheri99, Werner99, Greentree00}. Since neither of these linewidths is an essential part of the emitter/field interaction this work left open the fundamental question as to whether ensembles could generate single-photon nonlinearities.   

In 2006 Hartmann, Brandao, and Plenio (HBP) accomplished an exact quantum treatment of the generation of an optical Kerr nonlinearity by an ensemble~\cite{Hartmann06, Abadie11}. They did so by removing the complication caused by the atomic and cavity damping, considering the symmetric subspace of the ensemble (appropriate so long as the atomic damping is sufficiently small), and performing a Dyson expansion to determine the perturbative effect of the field on this subspace. They obtained a very different result than that derived in the semi-classical regime; due to the collective coupling to the ensemble, the coupling rate between the ensemble ground state and the first collective excited state is given by  $\tilde{g} = \sqrt{N}g$, where $N$ is the number of emitters and $g$ is the coupling for a single emitter. Since the perturbative regime requires that $g$ is much less than the detuning, the maximum allowed value for $g$ reduces as $1/\sqrt{N}$. This reduction of $g$ exactly balances the increase in the Kerr nonlinearity as the number of emitters is increased. The maximum rate of the nonlinearity generated by the ensemble is thus no more than that which can be generated by a single emitter. 

While the 2006 paper by HPB is famous, the results discussed above do not appear to be widely known~\cite{Venkataraman13, Trivedi19}. In 2013 Venkataraman, Saha, and Gaeta (VSG) performed an experiment in which they realized a Kerr nonlinearity using an ensemble with an average of 20 photons, and showed that their results were consistent with the  standard semiclassical analysis. In 2019 Trivedi~\textit{et al.\ }considered the generation of nonlinearities by an ensemble in a cavity at the single-photon level using a recently developed numerical technique~\cite{Trivedi19}. While their results were not inconsistent with those of HBP, they did not determine how the nonlinearity scales with the size of the ensemble.   

Here we introduce a comprehensive, fully quantum mechanical theory for the generation of nonlinearities by single emitters and ensembles of identical independent emitters. This theory provides insight into the physics underlying the generation of nonlinearities and allows us to immediately answer a number of the outstanding questions. We expect that the theory will provide a crucial foundation for answering important questions that remain.  As we will show, our theory also furnishes a powerful method for calculating the nonlinearities generated by emitters, and thus for designing driving schemes to engineer nonlinear processes. 

In addition to the theory itself and the associated methods, our primary results are as follows. First, we confirm that in the few-photon regime the nonlinearities generated by ensembles are severely limited by the collective coupling, and we are able to show how the very different semi-classical behavior emerges when the fields are in coherent states with sufficiently high amplitude. Second, we show surprisingly that ensembles of independent emitters \textit{can} generate nonlinearities that increase in strength with the number of emitters. Outside the rotating-wave regime (when the emitter/field coupling is sufficiently strong, but still small compared to the transition frequencies) the transition frequencies themselves take the place of the detunings in the perturbation theory. In this case, the emitters effectively act independently, the nonlinearities scale with the number of emitters, and very large  nonlinearities can be generated. The mechanism that enables this scaling is essentially the same as that which does so in the semi-classical regime. 

Given the above results, it is clear that the ability of ensembles to generate giant nonlinearities at low photon numbers in the rotating wave regime will depend on precisely \textit{where} (at what photon number) the transition to  semi-classical behavior occurs. We do not resolve this question here, but we expect that it can be answered with the tools we have developed. 

\subsection*{Structure of this article}

In Section~\ref{sec2A} we introduce the Hamiltonian for an ensemble of independent multilevel emitters interacting with one or more cavity modes, as well as the form this Hamiltonian takes under the rotating-wave approximation. In Sections~\ref{sec2B} and~\ref{sec2C} we briefly review multi-parameter time-independent perturbation theory (TIPT), present the recursion relations that determine the expansion coefficients, and introduce some useful short-hand notation. In Section~\ref{sec2D} we show how TIPT can be extended to perturbations that involve operators of an external subsystem. We then show, in Section~\ref{SecHeff}, that the expansion for the perturbed ground-state ``eigenvalue" in this extended version of TIPT gives precisely the effective Hamiltonian for the external perturbing system. In our case it is the field mode(s) that are perturbing the emitter and for which the effective nonlinear Hamiltonian is induced. In Sections~\ref{Sec2Eb} and~\ref{Sec2F} we discuss, respectively, limits on the atom/field coupling rates and the frequency-matching conditions for the generated nonlinearities. In Section~\ref{sec2G} we show how to derive the full effective master equation for the field; due to decay of the emitter levels the emitter induces dissipation for the field in addition to the effective nonlinear Hamiltonian. In Section~\ref{sec2H} we elucidate the fact that the initial state of the emitter and field is important in generating the nonlinearities, and this restricts the speed at which the field can be changed. In Section~\ref{secSI}, as an example we use our method to calculate the cross-Kerr nonlinearity generated by the Schmidt-Imamo{\=g}lu scheme employing a single 4-level emitter. With that we complete our development of the theory of the generation of nonlinearities by a single emitter. 

In Section~\ref{secEns} we are finally ready for our ultimate goal, that of calculating the nonlinearities generated by ensembles. In Section~\ref{tlsym}, to provide insight, we use our method to do this for the simplest example, an ensemble of undriven two-level systems.  In Section~\ref{Arbsym} we show how to apply the method developed for single emitters to ensembles of emitters. We then use it to show in general how the self-Kerr nonlinearity generated by an ensemble relates to that generated by a single emitter. Next we turn to the pivotal question, that of how the bound on the coupling rates scales with the size of the ensemble? In Section~\ref{barecoup} we answer this for ensembles in the bare coupling regime, and in Section~\ref{RWAbound} for the RWA regime. In Section~\ref{KerrScale} we show that in the RWA regime the scaling of the Kerr nonlinearity for two-level systems is quite different to that for the Schmidt-Imamo{\=g}lu scheme. In Section~\ref{semiclass} we show that when the field modes are in coherent states with sufficiently high amplitudes the size of the nonlinearities scales linearly with the size of the ensemble, thus explaining the emergence of this behavior in the semi-classical regime. In  Section~\ref{travwave} we show how to apply our method to calculate nonlinearities for travelling-wave fields and write these as nonlinear susceptabilities. In Section~\ref{potstr} we point out that the ability of ensembles to generate nonlinearities for weak fields will depend on exactly where the emitter/field system makes the transition to the semiclassical regime. While we do not explore this question further here, we discuss some recent experimental results in this context.  Section~\ref{conc} concludes with a discussion of open questions. 

\section{Generation of Nonlinearities by a single emitter}
\label{sec2}

\subsection{A Multilevel emitter coupled to field modes} 
\label{sec2A}  

A general emitter has a set of discrete states, $\ket{\tilde{n}}$, $n=0, 1, 2, \ldots$, with energies $\tilde{\mathcal{E}}_n$. By convention $\ket{\tilde{0}}$ is the ground state. Transitions between emitter states are induced by exchanging energy with the field. While all field modes are coupled to all emitter transitions, only coupling between modes and transitions that are sufficiently close in energy need to be included. For our purposes we need only to have each field mode coupled to a single transition, although the method we introduce can certainly be applied in the general case. If we have $L$ modes with mode operators $a_l$, $l = 1, \ldots, L$, each of which is coupled to a transition $\ket{\tilde{n}_l} \leftrightarrow \ket{\tilde{k}_l}$, under the usual dipole approximation the Hamiltonian is~\cite{Tannoudji89} 
\begin{align}
    H = \tilde{H}_0 + \sum_l  \hbar ( g_{l} a_l  + g_{l}^* a_l^\dagger  ) (\sigma_{l} + \sigma_{l}^\dagger) +  H_{\ms{f}}
    \label{H1}
\end{align}
with 
\begin{align}
  \tilde{H}_0 & = \sum_{\tilde{n}} \tilde{\mathcal{E}}_n \ket{\tilde{n}}\bra{\tilde{n}}, \\
  H_{\ms{f}}  & = \sum_l \hbar \omega_l a_l^\dagger a_l \\ 
  \sigma_l & =  \ket{\tilde{n}_l}\bra{\tilde{k}_l}, 
\end{align} 
in which $\sigma_l$ is called the \textit{transition operator} for the transition $\ket{\tilde{n}_l} \leftrightarrow \ket{\tilde{k}_l}$. We use the convention that $\sigma_l$ always denotes a lowering operator, meaning that the energy of $\ket{\tilde{k}_l}$ is greater than that of $\ket{\tilde{n}_l}$. The reason that we use tildes on some operators and states will become clear below. In short, tildes denote the original emitter Hamiltonian and its eigenbasis and will be removed to denote the emitter Hamiltonian that includes classical driving terms and its eigenbasis. 

If a mode coupled to a transition is in a coherent state with amplitude $\alpha$, in which $|\alpha|^2 \gg 1$, then the mode operators $a$ and $a^\dagger$ can be replaced by $\alpha$ and $\alpha^*$ (the reason for this is detailed in Section~\ref{semiclass}) so that the mode is eliminated from the dynamics. This results in a coupling between the upper and lower states proportional to $\alpha$ and is referred to as a \textit{classical drive}. Here we are interested in the situation in which classical drives are applied to various transitions of an emitter while other transitions may be coupled to modes as above. Alowing the first $M$ modes to be classical drives with amplitude $\alpha_l$, and defining the Hermitian operators 
\begin{align}
   \hat{x}_j = \sigma_{j} + \sigma_{j}^\dagger ,  
\end{align}  
this more general situation is described by the Hamiltonian 
\begin{align} 
    \tilde{H}_{\ms{D}} & = \tilde{H}_0 + \sum_{l = 1}^M  \hbar D_l \hat{x}_l +   \sum_{l=M+1}^{L} \hbar   ( g_{l} a_l + g_{l}^* a_l^\dagger  )   \hat{x}_{l} + H_{\ms{f}}  ,  \label{HD}
\end{align}
where $D_l = g_{l} \alpha_l  + g_{l}^* \alpha_l^*$.

To write $\tilde{H}_{\ms{D}}$ in the form appropriate for treating the interaction with the field modes as a  perturbation, we now diagonalize the emitter part of the Hamiltonian using a unitary change of basis, $U$, so that $\tilde{H}_{\ms{D}}$ becomes 
\begin{align}
    H_{\ms{B}} = H_0 +  \sum_{l=0}^{M-1} \hbar  ( g_{l} a_l + g_{l}^* a_l^\dagger  )   \Lambda_{m_l} + H_{\ms{f}} , 
    \label{HA}
\end{align} 
where 
\begin{align}
    H_0 & = \sum_{n} E_{0}^{(n)} \ket{n_0}\bra{n_0}  =  U^\dagger \left[ \tilde{H}_0 + \sum_{m \in \mathcal{S}_{\mbox{\tiny c}}}  \hbar D_m \hat{x}_m \right] U  \label{defUB} \\
    \Lambda_{m_l} & =  U^\dagger \hat{x}_{m_l}  U 
\end{align}
and $U$ is defined by Eq.(\ref{defUB}). The reason that we have labelled the above Hamiltonian with the subscript ``B" will be explained below. 

While $H_{\ms{B}}$ describes a driven emitter interacting with a number of field modes, it is not the Hamiltonian usually considered for optical emitters. For such emitters the transition frequencies are very much larger than the Rabi frequencies or coupling rates to the optical modes. In this case, so long as the frequencies of the classical drives and quantum modes satisfy a ``consistency" condition (see below) it is possible to move into the interaction picture so that the emitter Hamiltonian that remains contains the detunings between the fields and the transitions rather than the energies of the levels. Making the rotating-wave approximation then eliminates the time dependence in the interaction Hamiltonian. 

To implement the above procedure we return to the undiagonalized version of the emitter/field Hamiltonian, Eq.(\ref{HD}). We also need to determine the reference Hamiltonian for the emitter, $H_{\ms{Ref}}$, to move into the interaction picture with respect to. To proceed we note that the time dependence induced in the transition operator $\sigma_l$ and field operator $a_l$ by moving to the interaction picture w.r.t an emitter Hamiltonian $\tilde{H}_{\ms{Ref}} \equiv \sum_{\tilde{n}} \mathcal{E}_n \ket{\tilde{n}}\bra{\tilde{n}}$ and the field Hamiltonian $H_{\ms{f}}$ are 
\begin{align}
    \sigma_l(t) & = \sigma_l \exp[-i\eta_l t], \;\;\;  \eta_l = (\mathcal{E}_{n_l} - \mathcal{E}_{k_l})/\hbar , \\ 
    a_l(t) & = a_l \exp[-i\omega_l t] . 
\end{align}
We need to choose the energies  $\mathcal{E}_n$ so that the frequency of each transition, $\eta_l$, is equal to that of the mode to which it is coupled,  $\omega_l$. This is not always possible, but if it is we will refer to the set of mode frequencies $\{\omega_l\}$ as \textit{consistent}. 

%\todo{minor remark on order: we introduce a diagonalized Hamiltonian above but now we are talking about the initial not-diagonalized form}
Moving into the interaction picture w.r.t $H_{\ms{Ref}}$ and $H_{\ms{f}}$, and assuming that the set of mode frequencies is consistent, the interaction picture Hamiltonian is 
\begin{align}
  \tilde{H}_{\ms{I}} & = \Delta \tilde{H} + \sum_l  \hbar ( g_{l} a_l e^{-i\omega_l} + \mbox{H.c.}  ) (\sigma_{l} e^{-i\omega_l}  + \mbox{H.c.} )   \nonumber \\
  & = \Delta \tilde{H} + \sum_l  \hbar ( g_{l} a_l \sigma_{l}^\dagger +  g_{l}^* a_l^\dagger \sigma_{l} ) \nonumber \\ 
  & \;\;\; + \sum_l  \hbar ( g_{l} a_l \sigma_{l} e^{-i2\omega_l}  +  g_{l}^* a_l^\dagger \sigma_{l}^\dagger e^{i2\omega_l}  ) 
\end{align} 
with 
\begin{align}
  \Delta \tilde{H} & = \sum_{\tilde{n}} \Delta\mathcal{E}_n \ket{\tilde{n}}\bra{\tilde{n}} = \sum_{\tilde{n}} (\tilde{\mathcal{E}}_n - \mathcal{E}_n) \ket{\tilde{n}}\bra{\tilde{n}} . 
\end{align}
The last term in $\tilde{H}_{\ms{I}}$ is time-dependent. Since the transition frequencies are much larger than all other timescales in the dynamics (the interaction rates $g_l$ and the detunings $\Delta\mathcal{E}_n$), the rapidly oscillating terms all but cancel themselves out and contribute little to the evolution. The ``rotating-wave" approximation involves discarding these terms. 

We now allow some of the field modes to contain large coherent states and thus provide classical driving as before. Choosing the first $M$ modes to be the classical drives the emitter-field Hamiltonian is  
\begin{align}
  \tilde{H}_{\ms{I}} & = \Delta \tilde{H} + \sum_{l=0}^{M}  \hbar ( \beta_{l} \sigma_{l}^\dagger +  \mbox{H.c.} ) + \sum_{l=M+1}^{L}  \hbar ( g_{l} a_{l} \sigma_{l}^\dagger + \mbox{H.c.} ) ,  
\end{align} 
in which $\beta_l = g_l \alpha_l$. The Hamiltonian $\tilde{H}_{\ms{I}}$ is the large-transition-frequency version of $\tilde{H}_{\ms{D}}$ above, in which an arbitrary emitter is driven by classical fields in a time-independent way and interacts with a number of quantum mechanical field modes. This Hamiltonian describes all current schemes for the generation of optical nonlinearities by driven emitters. 

There are two differences between $\tilde{H}_{\ms{D}}$ and $\tilde{H}_{\ms{I}}$. In $\tilde{H}_{\ms{I}}$ the energy levels of the emitter are replaced by the detunings, $\Delta\mathcal{E}_n$, between the driving fields (or modes) and the emitter transitions. The second difference is that the interaction between the emitter and the field modes has a different form. This will turn out to have a profound effect on the behavior we study here. Note that for the classical driving to have the time-independent form in $\tilde{H}_{\ms{I}}$ the frequencies of these driving fields must be \textit{consistent} in the manner defined above. Only if they are can we treat the driving as being time-independent in the frame of the interaction picture. On the other hand, to treat the driving as time-independent in the Schroedinger picture ($\tilde{H}_{\ms{D}}$) all we need is that the driving actually be time-independent (meaning that the (complex) amplitudes of the driving fields are constant). 

To put the Hamiltonian $\tilde{H}_{\ms{I}}$ in the form appropriate for treating the interaction with the field modes as a perturbation, we now diagonalize the emitter part of the Hamiltonian using a unitary transformation $U$. The result is 
\begin{align}
    H_{\ms{R}} & = \Delta H + \sum_l K_l^\dagger a_l + K_l a_l^\dagger  
    \label{baseH2}
\end{align}
in which 
\begin{align}
    \Delta H & = \sum_{n} E_{0}^{(n)} \ket{n_0}\bra{n_0} \nonumber \\ 
    & = U^\dagger \left[ \Delta \tilde{H} + \sum_l  \hbar ( \beta_{l} \sigma_{l}^\dagger +  \beta_{l}^* \sigma_{l} ) \right] U \label{defU2} \\
    K_l & = g_l U^\dagger \sigma_{l} U 
\end{align}
and $U$ is defined by Eq.(\ref{defU2}). 

We now have two different Hamiltonians that describe the interaction of multi-level emitters with the electromagnetic field. In the first, $H_{\ms{B}}$, given in Eq.(\ref{HA}), the energy levels for the emitter are the emitter's actual, or ``bare" energy levels, which is the origin of the choice of subscript ``B". In the second, $H_{\ms{R}}$, given in Eq.(\ref{baseH2}), these energy levels are ``relative" to those of the driving fields, which is the reason for the subscript ``R". 

The regime in which an emitter will generate a distinct set of nonlinearities for the field modes is that in which the size of the coupling with each of the field modes is much smaller than the energy separation (the relative separation in the case of $H_{\ms{R}}$) between the two levels of the transitions to which they couple. That is, when 
\begin{align}
    |g_l| & \ll E_0^{(n_l)} - E_0^{(k_l)} . 
\end{align}
In this case, the interaction with the field is a perturbation on the dynamics of the emitter. We will use time-independent perturbation theory to determine the resulting dynamics of the field modes.

\subsection{Time-independent perturbation theory (TIPT)}
\label{sec2B}

% We will use multi-parameter time-independent perturbation theory (" TIPT") to determine the nonlinearities generated by a  coherently driven multi-level emitter. The regim where emitters generate nonlinearities for a set of field modes is that in which field modes are off-resonant with the transitions to which they couple:  must be larger than the Rabi frequency that the mode would induce for the emitter transition if the detuning were zero and the mode were in a coherent state containing on average one photon. When a field mode is resonant  a transition the coupling induces energy exchange between the field and the emitter. As will become clear, When the mode is far enough off-resonance the effect on the emitter dynamics is small, and the field(s) undergo an evolution described by a Hamiltonian that can be written as a power series in the field coupling operator. The terms in this power series r the various nonlinearities generated by the emitter.

% A key innovation of our approach is to employ (multi-parameter) time-independent perturbation theory (TIPT). 

Recall that time-independent perturbation theory (hereafter TIPT) allows one to calculate the eigenvalues and eigenvectors of a Hamiltonian 
\begin{align}
    H = H_0 + \lambda V 
\end{align} 
in terms of the eigenvectors and eigenvalues of $H_0$ and as a power series expansion in the parameter $\lambda$. For this expansion to be valid the magnitude of the off-diagonal elements of $\lambda V$ in the basis of $H_0$ must be smaller than the separation between the corresponding energy levels of $H_0$. If the size of the energy level separations are on the order of $\Delta$, and the size of the elements of $V_l$ are on the order of $v$, then the power series that results is actually a power series in the small parameter $\lambda v/\Delta$. For future reference it is useful to define the following two terms: 

\begin{itemize}
    \item \textit{Expansion parameter:} This is a real parameter that multiplies a perturbative term in the Hamiltonian. For $H$ above, $\lambda$ is an expansion parameter.
    \item \textit{Perturbation parameter:} This is a real parameter that gives the relative size of the perturbation with respect to the base Hamiltonian $H_0$. One usualy assumes that this parameter is smaller than unity because if it is not the  perturbation expansion is not guaranteed to be valid. The perturbation parameter is typically on the order of $\lambda |V|/|H_0|$, where $|\cdot|$ denotes an operator norm. For a given expansion parameter, $\lambda$, we will denote the corresponding perturbation parameter by $\epsilon_\lambda$. 
\end{itemize}

To determine the power series for the eigenvectors and eigenvalues of $H = H_0 + \lambda V$ one assumes that they may be written in the form 
\begin{align}
    E^{(n)} & = E^{(n)}_0 + \sum_{j\geq 1} E^{(n)}_j \lambda^j  , \\
    |n\rangle & = |n_0\rangle + \sum_{j\geq 1} \lambda^j |n_j\rangle , 
    \label{nlambda}
\end{align}
where $E^{(n)}_0$ and $|n_0\rangle$ are the eigenvectors and eigenvalues of $H_0$. The expansion coefficients $E^{(n)}_j$ and $|n_j\rangle$ are obtained from the equation $H \ket{n} = E^{(n)}\ket{n}$ by multiplying on the left by the eigenstates of $H_0$ and equating the coefficients of the powers of $\lambda$~\cite{Jacobs14}. This procedure provides recursion relations for the expansion coefficients from which explicit expressions can be obtained. 

% Here, we will need to employ TIPT with more than one perturbation operator, and a separate expansion parameter for each. The result is referred to as \textit{multi-parameter} TIPT and is a straightforward extension of the usual TIPT~\cite{Parusinski20}. If we have the following Hamiltonian with four perturbations,  
% \begin{align}
%     H & = H_0 + \lambda V  + \nu X  + \eta Y  + \xi Z   
%     \label{H4p}
% \end{align}
% where $\lambda$, $\nu$, $\eta$, and $\xi$ are the four expansion parameters, then the expansions for the eigenvalues and eigvectors are  
% \begin{align}
%     E^{(n)} & = \sum_{j,k,l,q} E^{(n)}_{jklq} \lambda^j \nu^k \eta^l \xi^q , \label{Enx}\\
%     |n\rangle & = \sum_{j,k,l,q} \lambda^j \nu^k \eta^l \xi^q  |n_{jklq}\rangle . \label{Vnx}
% \end{align} 
% Here the sums over all the indices $j$, $k$, $l$, $q$, run from zero to infinity, and $E^{(n)}_{0000} = E^{(n)}_{0}$ and $|n_{0000}\rangle = |n_0\rangle$ are, respectively, the eigenvalues and eigenvectors of $H_0$. 

Here, since we want to consider an emitter interacting with up to four electromagnetic modes (so as to explicitly include the generation of third-order nonlinearities) we will need to calculate the eigenvalues and eigenvectors of a Hamiltonian $H$ with four perturbative terms:  
\begin{align}
H = H_0 + \lambda V + \nu X + \eta Y + \xi Z .  
\label{H4p}
\end{align}
Here $H_0$ is a Hamiltonian whose eigenvectors and eigenvalues are known and $\lambda$, $\nu$, $\eta$, and $\xi$ are expansion parameters associated with the perturbative terms $V$, $X$, $Y$, and $Z$, respectively. Looking ahead briefly, these perturbative Hamiltonians will have the form 
\begin{align}
    V & = A^\dagger a + A a^\dagger , \;\;\;\; X = B^\dagger b + B b^\dagger \label{Vform1} \\
    Y & = C^\dagger c + C c^\dagger , \;\;\;\; Z = D^\dagger d + D d^\dagger , \label{Vform2}   
\end{align}
in which $a$, $b$, $c$, and $d$ are mode operators for four different electromagnetic modes, and the operators $A$, $B$, $C$, and $D$ will in general be lowering operators (or sums of more than one lowering operator) each of which couple two levels of the emitter.  

To compute the eigenvalues and eigenvectors of $H$, we use multi-parameter time-independent perturbation theory (TIPT). This method extends the usual TIPT to the case where multiple perturbations are present~\cite{Parusinski20}. To do so we assume that the eigenvalues and eigenvectors of $H$ may be written in the form 
\begin{align}
    E^{(n)} & = \sum_{j,k,l,q} E^{(n)}_{jklq} \lambda^j \nu^k \eta^l \xi^q , \label{Enx}\\
    |n\rangle & = \sum_{j,k,l,q} \lambda^j \nu^k \eta^l \xi^q  |n_{jklq}\rangle , \label{Vnx}
\end{align} 
where the sums over all the indices $j$, $k$, $l$, $q$, run from zero to infinity, and $E^{(n)}_{0000} = E^{(n)}_{0}$ and $|n_{0000}\rangle = |n_0\rangle$ are, respectively, the eigenvalues and eigenvectors of $H_0$. To determine the eigenvectors and eigenvlues of $H$ we use the eigenvalue equation $H \ket{n} = E^{(n)}\ket{n}$ and substitue in the expansion forms for $E^{(n)}$ and $|n\rangle$ given above. As in single-paramter TIPT we derive recursion relations for the expansion coefficients $E^{(n)}_{jklq}$ and $|n_{jklq}\rangle$ by multiplying the eigenvalue equation on the left by $\bra{m_0}$. Solving the recursion relations we obtain expressions for the expansion coefficients (see Table~\ref{tab1}). %\todo{maybe a column specifying the resonance conditions; maybe more detailed explanation of the form column} 
We then use the eigenvalue and eigenvector expansions to calculate quantities of interest for the system described by the Hamiltonian $H$.

When using TIPT we can either choose the expansion parameters to be interaction rates between the primary and the perturbing systems, or we can include all physical content in the perturbation operators, $V$, $X$, \textit{etc}., relegating the expansion parameters to a purely formal role as dimensionless quantities. If we choose the latter, which is conventional in some circles, then we set the expansion parameters to unity in the above expressions for the eigenvectors and eigenvalues, so that 
\begin{align}
    E^{(n)} & = \sum_{j,k,l,q} E^{(n)}_{jklq}  , \\
    |n\rangle & = \sum_{j,k,l,q}  |n_{jklq}\rangle . 
\end{align} 
In the example we treat in Section~\ref{secSI}, we choose the expansion parameters to be interaction rates and thus use Eqs.(\ref{Enx}) and (\ref{Vnx}).
%\todo{why did we split the expansion parameter and the perturbation parameter; e.g., why include all physical content in the perturbation operator} 

\subsection{Calculating the TIPT expansion coefficients}
\label{sec2C}

As noted above, recursion relations for the TIPT expansion coefficients can be derived from the eigenvalue equation for $H$. The recursion relations for a three-parameter expansion, in which the perturbation Hamiltonians are $V$, $X$, and $Y$, are  
\begin{widetext}
\begin{align}
    E^{(n)}_{jkl} & = \sum_{q\not=n}\, \Bigl[  V_{nq} \langle q_0 |n_{j-1,k,l} \rangle +   X_{nq} \langle q_0 | n_{j,k-1,l} \rangle +   Y_{nq} \langle q_0 | n_{j,k,l-1} \rangle \Bigr] \, - \!\! \sum_{\substack{q,p,r \not= 000 \\ q,p,r \not= jkl}}^{j,k,l} E^{(n)}_{qpr} \langle n_0 | n_{j-q,k-p,l-r} \rangle 
\end{align} 
\begin{align}
    \langle m_0 | n_{j,k,l} \rangle  & = \frac{1}{ \Delta_{nm}} \sum_{q\not=m}\, \Bigl[ V_{mq} \langle q_0 |n_{j-1,k,l} \rangle + X_{mq} \langle q_0 | n_{j,k-1,l} \rangle +  Y_{mq} \langle q_0 | n_{j,k,l-1} \rangle \Bigr] \, - \!\!  \sum_{\substack{q,p,r\not= 000 \\ q,p,r\not= jkl}}^{j,k,l} \Biggl( \frac{E^{(n)}_{qpr}}{ \Delta_{nm}}\Biggr)\, \langle m_0 | n_{j-p,k-q,l-r} \rangle , \;\;\; m \not= n , 
\end{align}
\begin{align}
    \langle n_0 | n_{j,k,l} \rangle & = - \frac{1}{2} \sum_{\substack{q,p,r\not= 000 \\ q,p,r \not= jkl}}^{j,k,l} \sum_{m} \langle n_{q,p,r} |m_0 \rangle \langle m_0 |  n_{j-q,k-p,l-r} \rangle   
\end{align}
\end{widetext} 
where $|n_{000}\rangle \equiv |n_0\rangle$ and we have defined 
\begin{align}
  \Delta_{nm} \equiv E^{(n)}_0 - E^{(m)}_0 .
\end{align} 
The summation subscript $x,y,z \not= abc$ means that the single term in which $x=a, y=b, z=c$ is excluded from the sum. If not specified, all sums run from $0$ though $N-1$ where $N$ is the dimension of the emitter. We always set 
\begin{align}
     V_{nn} & = X_{nn} = Y_{nn} = 0  
\end{align}
since any non-zero diagonal elements of the perturbation operators can always be absorbed into $H_0$. A result of this convention is that 
\begin{align}
 E^{(n)}_{100} & = E^{(n)}_{010} = E^{(n)}_{001} = 0 \\ 
 \langle n_0 |n_{100}\rangle & = \langle n_0 |n_{010} \rangle = \langle n_0 |n_{001} \rangle = 0 .
\end{align}
%The generalization to recursion relations for TIPT expansions for more than three parameters is clear from the three parameter relations above. 
We give the recursion relations for three instead of four parameters here because i) the four-parameter recursion relations do not fit neatly on the page, and ii) given the three parameter relations it is simple to generalize to any number. 

There is a great deal of symmetry in the expansion coefficients. For example, the expansion coefficient $E^{(n)}_{010}$ is the same as $E^{(n)}_{100}$ but with $V$ replaced by $X$. Similarly, the 2-parameter expansion coefficient $E^{(n)}_{jk}$ is the same as the 4-parameter expansion coefficient $E^{(n)}_{jk00}$, and is the same as $E^{(n)}_{0j0k}$ up to the replacements $V \rightarrow X$ and $X \rightarrow Z$. Thus specifying the expansion coefficient $E^{(n)}_{jk}$ gives us all expansion coefficients for which exactly two indices are nonzero.  

We now introduce a compact notation that will help to present all distinct coefficients for expansion terms to $4^{\msi{th}}$ order for up to four fields (Table~\ref{tab1}). Consider a product of a set of symbols $A$, $B$, $C, \ldots$ in which each symbol has two subscripts. An example is the product $A_{ab} B_{bc} C_{cd}$. We define $\mathsf{ALLP}\mbox{\textbf{[}}\!\mbox{[} \, A_{ab} B_{bc} C_{cd} \cdots \,  \mbox{]}\!\mbox{\textbf{]}}$ as the sum of products of all permutations of the symbols $A$, $B$, $C, \ldots$, where the subscripts do not permute with their symbols (the subscripts stay in the same place). The following examples provide clarification of these rules: 
\begin{align}
    \mathsf{ALLP}\mbox{\textbf{[}}\!\mbox{[} \, A_{ab} B_{bc} C_{cd}  \,  \mbox{]}\!\mbox{\textbf{]}} 
    & = A_{ab} B_{bc} C_{cd} + A_{ab} C_{bc} B_{cd} \nonumber\\
    & \;\;\; + B_{ab} A_{bc} C_{cd} + B_{ab} C_{bc} A_{cd} \nonumber\\
    & \;\;\; + C_{ab} A_{bc} B_{cd} + C_{ab} B_{bc} A_{cd} \\
    \mathsf{ALLP}\mbox{\textbf{[}}\!\mbox{[} \, A_{ab} B_{bc} B_{cd}  \,  \mbox{]}\!\mbox{\textbf{]}} 
    & = A_{ab} B_{bc} B_{cd}  + B_{ab} A_{bc} B_{cd} \nonumber\\
    & \;\;\; + B_{ab} B_{bc} A_{cd} . 
\end{align} 
Using this notation we give all distinct expressions for the expansion coefficients up to $4^{\msi{th}}$ order in Table~\ref{tab1}. Since we write these explicitly for the ground state ($n=0$), we use  
\begin{align}
    \Delta_{k} \equiv E^{(0)}_0 - E^{(k)}_0 . 
\end{align}
All coefficients for the ground state for up to four fields are obtained by taking the coefficients in Table~\ref{tab1} and permuting their subscripts, and/or setting one or more of the subscripts to zero.

\subsection{Applying TIPT to perturbations by external systems} 
% [2023.01.04 - de]
% In this section, we will extend time-independent perturbation theory (TIPT) to the case where the perturbation is due to an interaction with another system...Consider a single mode and a set of emitter states. For each emitter state, there is a subspace consisting of every state of the mode. If the energy separations between the mode states << those of the emitter and the mode is not too highly populated, the subspaces corresponding to each emitter state r  separated in energy....

\label{sec2D}

\begin{figure}[t]
\centering
\leavevmode\includegraphics[width = 0.75 \columnwidth]{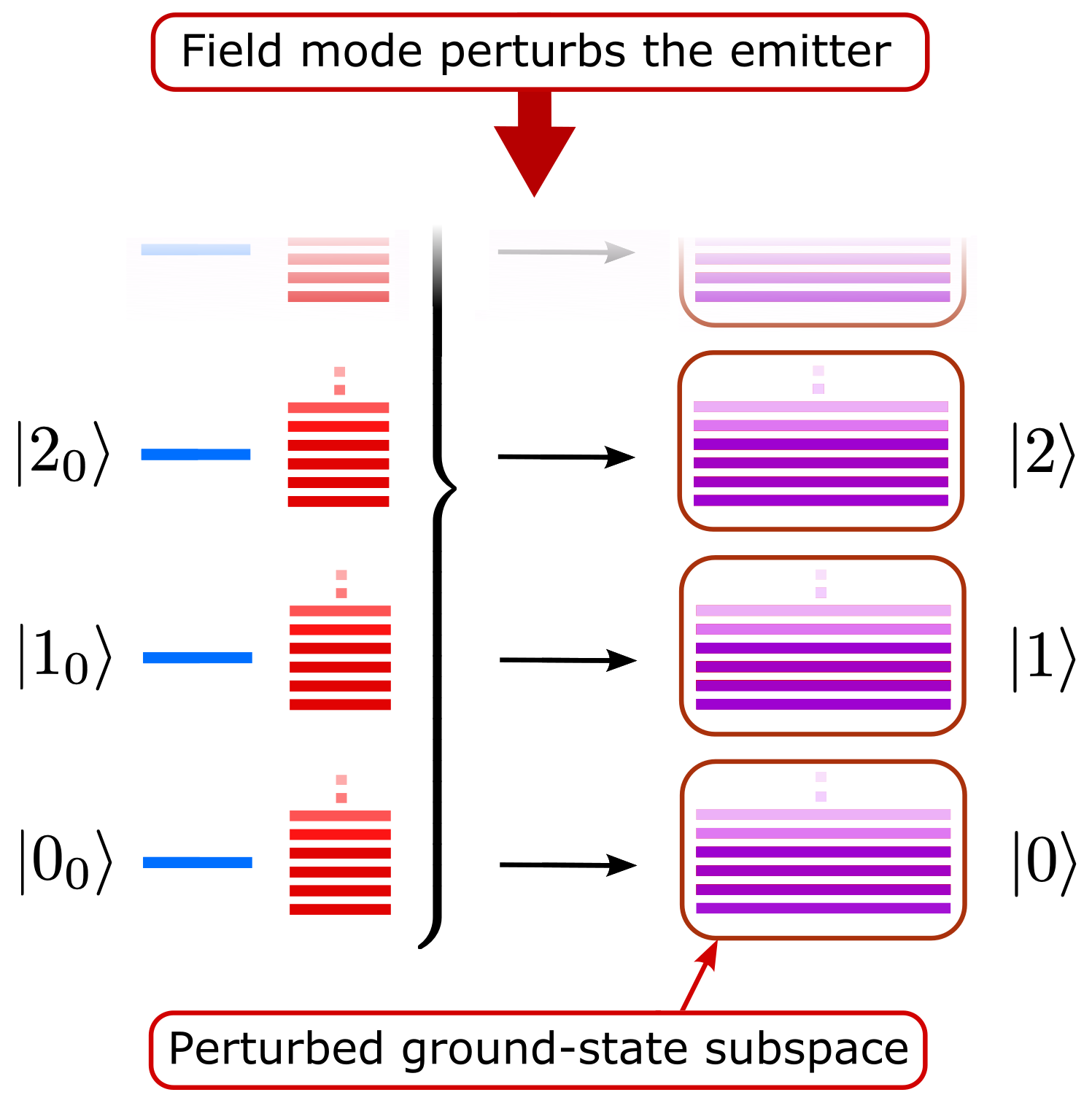}
\caption{(Color online) Here the emitter states, denoted by $\ket{0_0}, \ket{1_0}, \ldots$, are depicted in blue and the field states are depicted in red. The emitter interacts with the field mode where the interaction is perturbative from the point of view of the emitter. When the systems are uncoupled each emitter state defines a subspace of the joint system. The TIPT perturbation expansion determines new subspaces, each a perturbation of these original subspaces, on which the joint Hamiltonian is block diagonal. The perturbed emitter ``states" $\ket{n}$ are shorthand for ``any state in the perturbed subspace corresponding originally to $\ket{n_0}$". The ``eigenvalues'' calculated by TIPT are now operators, $\hat{E}^{(n)}$, and give the action of the Hamiltonian on each of the perturbed subspaces. Each operator $\hat{E}^{(n)}$ preserves the subspace $\ket{n}$ and is the effective nonlinear Hamiltonian for the field mode(s) on this subspace.} 
\label{fig:extpert} 
\end{figure}

\begin{figure*}
\centering
\leavevmode\includegraphics[width = 2.05 \columnwidth]{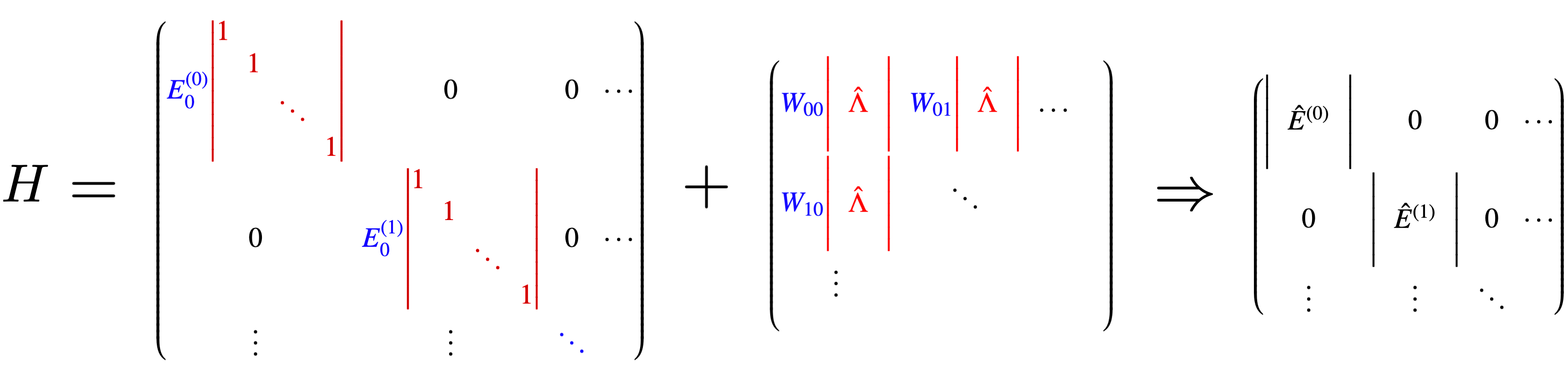} 
\caption{(Color online) The perturbation expansion block-diagonalizes the Hamiltonian. The joint Hamiltonian of the emitter and field has two terms, depicted here on the left hand side. The first matrix is a tensor product of the Hamiltonian of the emitter and the identity for the field. The second is the interaction. Here we have depicted the simplest case in which $V = W \otimes \Lambda $ in which $\Lambda$ is an operator of the field. In this case each of the elements $V_{ij} = W_{ij} \Lambda$. The elements of emitter operators are shown in blue and the field operators are shown in red. On the right hand side is the block-diagonal form in which the emitter and field basis states are now mixed together. The basis for the matrices on the left before the perturbation is $|n_0\rangle|j\rangle$, while on the right after the perturbation there is no neatly factored basis. Nonetheless, on the right each block corresponds to a subspace with a well-defined basis of polariton states, and which has a signifcant overlap with the subspace defined by $|n_0\rangle$. Consult the main text for how to complete the diagonalization and for a special case in which the perturbed basis can still be factored out.} 
\label{figblock} 
\end{figure*}

We now consider applying TIPT to the situation in which an interaction between a ``primary" system and another system is perturbative (small) compared to the energy separation between the states of the primary system but not compared to the energy separations of the states of the secondary system. In this case, rather than diagonalizing the Hamiltonian of the primary system alone, TIPT will allow us to \textit{block-diagonalize} the Hamiltonian. Each block corresponds to (is a perturbed version of) one of the subspaces defined by the original eigenstates of the primary system. This situation is depicted in Fig.~\ref{fig:extpert}. The unperturbed subspaces, each defined by a state of the primary system, are separated in energy, and applying TIPT with respect to the primary system dertermines corresponding perturbed subspaces that are not mixed together by the joint Hamiltonian. The joint Hamiltonian thus acts non-trivially only within the subspaces. We show the block-diagonalization of the Hamiltonian written as a matrix in Fig.\ref{figblock}. 

In our case the primary system will be an emiter and the secondary system (or systems) will be modes of the electromagnetic field. Thus the subspaces defined by the states of the emitter will be separated in energy so long as the modes are not too highly populated. (Note that for the rotating-wave interaction all the states of the modes are effectively degenerate, a degeneracy which is removed by the interaction.)

We will find that the action of the full Hamiltonian on each subspace is nonlinear for the field modes and is different on each subspace. It is this action that gives the effective nonlinearities for the fields. Note that since the eigenstates making up each subspace are weakly entangled ``polariton'' states of the emitter and the field modes, to generate the nonlinear evolution for the fields alone would require turning on the interaction adiabatically to transform the field states to the polariton states, allowing the effective nonlinearity to act, and then turning of the interaction adiabatically.  

We now turn to the mathematical details. Consider a single perturbation parameter in which the perturbation is due to an interaction with another system. In this case the Hamiltonian is 
\begin{align}
    H = H_0 \otimes I  + \lambda V 
\end{align}
in which $H_0$ is the Hamiltonian of the emitter, $I$ is the identiy operator for the perturbing system, and the interaction operator $V$ is in general a sum of tensor products of operators of the emitter and the perturbing system: 
\begin{align}
    V = \sum_j W^{j} \otimes \Lambda_j .  \label{wjlj}
\end{align}
Using the matrix representation of the tensor product, $V$ is now a matrix indexed by the states of the emitter in which each element, $V_{jk}$, is an operator that acts on the perturbing system (equivalently $V_{jk}$ is a matrix with the dimensions of the perturbing system):  
\begin{align}
     V = \left( \begin{array}{ccc}
        \hat{V}_{11}  & \hat{V}_{12}  &  \cdots \\
        \hat{V}_{21}  & \hat{V}_{22}  &  \cdots \\
       \vdots   & \vdots  &  \ddots 
     \end{array}  \right) . 
\end{align}
If the perturbation is merely the product of a single Hemitian operator of the emitter, $W$, and an operator for the perturbing system, $\hat\Lambda$, then the elements of $V$ are simply $\hat{V}_{jk} =  W_{jk} \hat\Lambda$ as depicted in Fig.\ref{figblock}. Time-independent perturbation theory, as derived above in Sections~\ref{sec2B} and~\ref{sec2C}, can be applied directly to the general situation given in Eq.(\ref{wjlj}) by taking the elements $V_{jk}$ to be operators instead of numbers. The recursion relations for TIPT remain valid because in deriving them we were careful to respect the order in which the matrix elements of $V$ are multiplied together. Similarly the expressions for the terms in the expansions for the eigenvalues (Table~\ref{tab1}) and eigenvectors (Appendix~\ref{AppEigvec}) are valid when the matrix elements $V_{jk}$ are operators. 

To understand the meaning of the expansion for the eigenvectors when the matrix elements of $V$ are operators we begin by writing out the first few terms in this expansion (these terms are given in Appendix~\ref{AppEigvec}): 
% first note that TIPT as we have described it in Section~\cite{sec2B} can be applied to the situation of a perturbation by a second system simply by replacing the parameter $\lambda$ with a Hermitian operator, $\Lambda$, of that system: 
% \begin{align}
%      H = H_0 + \Lambda V . 
% \end{align} 
% Note that the way we have written $H$ here is standard shorthand for $H =  H_0 \otimes I_{\ms{f}} + V \otimes \Lambda$, where $I_{\ms{f}}$ is the identity operator for the field mode. Since any operator of the second system commutes with all the operators of the primary system, all the expressions we have derived for the exapnsion coefficents are still valid. To see exactly what the TIPT expansion means when the perturbation is an external system, note that the terms in the expansions for the eigenvalues and eigenvectors are now operators on the space of the external system (in our case a field mode). By using the expressions for the vectors in the eigenvector expansion given in Appendix~\ref{AppEigvec}, we can write the first few terms in the expansion for the eigenvectors:
\begin{align} 
    |n\rangle  = \sum_{j} \lambda^j   |n_{j}\rangle & = 
 \left[ 1 - \frac{\lambda^2}{2} \sum_{l\not= n} \frac{V_{nl}V_{ln}}{ \Delta_{nl}^2} + \ldots \right]  |n_0\rangle  \nonumber \\ 
                 & \;\;\;\; +  \lambda \sum_{l\not= n} \left[  \frac{V_{ln}}{ \Delta_{nl}}   + \lambda    \sum_{q\not=l,n} \frac{V_{lq}V_{qn}}{ \Delta_{nl}\Delta_{nq}} + \ldots  \right] |l_0\rangle
    \label{VnxF}
\end{align} 
Since the matrix elements of $V$ are field-mode operators  they operate on the field states. As such, the above expressions need some explanation because we have not included any field states for these operators to act on. We can think of the state $\ket{n_0}$ (and similarly $\ket{l_0}$) in the expansion above as representing any of the tensor product states $|n_0\rangle\ket{j}$ where $\ket{j}$ is a field state. Thus $\ket{n_0}$ represents the entire subspace $\{ \ket{n_0}\ket{j} : j = 0, 1, \ldots, \infty \}$. If we specify a particular field state on the right-hand side of Eq.(\ref{VnxF}) by replacing $\ket{n_0}$ by $\ket{n_0}\ket{j}$, then the left-hand side is the corresponding perturbed state of the joint system. The state $|n\rangle$ on the left-hand side of Eq.(\ref{VnxF}) thus represents all the states in the perturbed subspace  corresponding to the unperturbed subspace $|n_0\rangle$. Denoting the states in the perturbed subspace $\ket{n}$ by $\ket{n,j}$ we have 
\begin{align} 
    |n,j\rangle & =  
 \left[ 1 - \frac{\lambda^2}{2} \sum_{l\not= n} \frac{V_{nl}V_{ln}}{ \Delta_{nl}^2} + \ldots \right]  |n_0\rangle\ket{j}  \nonumber \\ 
                 & \;\;\;\; +  \lambda \sum_{l\not= n} \left[  \frac{V_{ln}}{ \Delta_{nl}}   + \lambda    \sum_{q\not=l,n} \frac{V_{lq}V_{qn}}{ \Delta_{nl}\Delta_{nq}} + \ldots  \right] |l_0\rangle\ket{j} . 
    \label{VnxF2}
\end{align} 
We will use $\ket{n}\bra{n}$ to denote a projector onto the perturbed subspace represented by $\ket{n}$. By multiplying both sides of the above equation on the left by $\bra{m_0}\bra{k}$ we obtain the coefficients of the state $\ket{n,j}$ in the original product basis $\ket{m_0}\ket{k}$. Denoting the matrix elements of the mode operator $V_{nl}$ by $V_{nl}^{(kj)}$ the result is  
\begin{align} 
   \bra{m_0}\!\bra{k} \! n,j\rangle & =  \left\{ \begin{array}{ll}
       \displaystyle 1 - \frac{\lambda^2}{2} \sum_{q} \sum_{l\not= n} \frac{ V_{nl}^{(kq)}V_{ln}^{(qj)}}{ \Delta_{nl}^2} + \ldots ,  & m = n  \\
        \displaystyle \lambda \sum_{r} \left[ \frac{V_{mn}^{(rj)}}{ \Delta_{nl}}   + \lambda  \!\! \sum_{q\not=m,n} \!\! \frac{ V_{mq}^{(kr)}V_{qn}^{(rj)}}{ \Delta_{nl}\Delta_{nq}} + \ldots  \right] ,   &  m \not= n 
    \end{array}  \right. 
    \label{VnxF3} 
\end{align}

By construction the action of the Hamiltonian on the perturbed subspace defined by $|n\rangle$ is given by $E^{(n)}$:
\begin{align}
    H \ket{n} = \hat{E}^{(n)} \ket{n} ,   
\end{align}
which also preserves the subspace (the action of $H$ is block-diagonal). To remind ourselves that $\hat{E}^{(n)}$ is an operator on the field modes we have added a ``hat" to it. This operator therefore gives the effective Hamiltonian for the field for subspace $\ket{n}$, so that in general there is a different effective Hamiltonian for each subspace.  

% The TIPT expressions for the expansion coeeficients also remain valid when the interaction between the primary and secondary system(s) is not merely the product of Hermitian operators for each system. For example, under the rotating-wave approximation the interaction between the two systems has the form 
% \begin{align}
%     H_{\ms{int}} = \lambda V = \lambda ( A a^\dagger + A^\dagger a)
% \end{align}
% in which $A$ is an operator of the primary system and $a$ is an operator of the secondary system. In that case we need merely note that the matrix elements of the interation operator, $V_{jk}$, are operators of the secondary system. The expansion coefficients in this case are given simply by substituting these operator-valued matrix elements into the expressions given in Table~\ref{tab1}

Finally, we will find it useful to examine the special case in which the emitter/field interaction is simply a product of a Hermitian operator of the emitter, which we will denote here by $V$, and a Hermitian operator $\Lambda$ of the field so that 
\begin{align} 
    H =  H_0 \otimes I + V \otimes \Lambda . 
\end{align}
For each eigenstate of $\Lambda$, which we will denote by $\ket{\lambda}$, the Hamiltonian becomes  
\begin{align}
    H =  H_0 + \lambda V  ,  
\end{align}
which is simply the usual single-system perturbation in which the value of the expansion parameter is the eigenvalue of  $\Lambda$. In this case the states of the perturbed subspace originating from the emitter state $\ket{n_0}$ are the product states $\ket{n(\lambda)}\ket{\lambda}$ where $\ket{n(\lambda)}$ is the perturbed emitter state when the scalar expansion parameter is equal to $\lambda$ (Eq.(\ref{nlambda})). We will use this fact in Sections \ref{sec2H} and \ref{barecoup}.

\begin{table*}[t]
    \centering
    \begin{tabular}{l|c|c|l}
    \hline
    \textit{Name} & \textit{Example} & \textit{Matching condition} & \; \textit{Coefficient in our compact notation}  \\
    \hline
    & & & \\[-3.3mm] 
    Frequency shift $\,$  &
    $ a^\dagger a \; $  &  \mbox{none}   &
    $ \, {\displaystyle \hat{E}_{2}^{(0)} = \sum_{l\not= 0} \frac{\mathsf{ALLP}\mbox{\textbf{[}}\!\mbox{[} \, V_{0l}V_{l0}  \mbox{]}\!\mbox{\textbf{]}}}{\Delta_l} } $  
    \\ 
    & & & \\[-3.8mm]
    \hline
    & & & \\[-3.3mm] 
    Linear coupling $\,$  &
    $ a^\dagger b  \; $  & $ \omega_a - \omega_b  =  0 $ &
    $ \, {\displaystyle \hat{E}_{11}^{(0)} = \sum_{l\not= 0} \frac{\mathsf{ALLP}\mbox{\textbf{[}}\!\mbox{[} \, V_{0l}X_{l0}  \mbox{]}\!\mbox{\textbf{]}}}{\Delta_l}  } $  
    \\
    &\\[-3.8mm]
    \hline
    & & & \\[-3.3mm] 
    &
    $a^\dagger a^2 \, $ & $ \omega_a  =  0 $ &
    $ \, {\displaystyle \hat{E}_{3}^{(0)} = \sum_{k,l\not= 0} \frac{\mathsf{ALLP}\mbox{\textbf{[}}\!\mbox{[} \, V_{0k}V_{kl}V_{l0}  \mbox{]}\!\mbox{\textbf{]}}}{\Delta_k \Delta_l} } $ 
    \\ 
    &\\[-3.8mm]
    \hline
    & & & \\[-3.3mm] 
    Frequency doubling $\,$  &
    $ a^{\dagger 2}  b \; $ & $ 2\omega_a - \omega_b  =  0 $ &
    $ \, {\displaystyle \hat{E}_{21}^{(0)}  = \sum_{k,l\not= 0}  \frac{\mathsf{ALLP}\mbox{\textbf{[}}\!\mbox{[} \, V_{0k}V_{kl}X_{l0} \, \mbox{]}\!\mbox{\textbf{]}}}{\Delta_k \Delta_l} } $ 
    \\
    &\\[-3.8mm]
    \hline
    & & & \\[-3.3mm] 
    Three-wave mixing $\,$  &
    $ a^\dagger b c \; $ &  $\omega_a - \omega_b - \omega_c =  0 $ &
    $ \, {\displaystyle \hat{E}_{111}^{(0)}  = \sum_{k,l\not= 0}  \frac{\mathsf{ALLP}\mbox{\textbf{[}}\!\mbox{[} \, V_{0k}X_{kl}Y_{l0} \, \mbox{]}\!\mbox{\textbf{]}}}{\Delta_k \Delta_l}  } $
    \\
    &\\[-3.8mm]
    \hline
    & & & \\[-3.3mm] 
    Self-Kerr $\,$ &
    $ (a^\dagger a)^2 \, $ &   \mbox{none}   &
    $ \, {\displaystyle \hat{E}_{4}^{(0)}  = \sum_{\substack{l,k,q \not= 0 \\ q \not= k,l}}  \frac{\mathsf{ALLP}\mbox{\textbf{[}}\!\mbox{[} \, V_{0k}V_{kq}V_{ql}V_{l0} \,  \mbox{]}\!\mbox{\textbf{]}}}{\Delta_q \Delta_k \Delta_l} - \sum_{k,l\not= 0}  \frac{\mathsf{ALLP}\mbox{\textbf{[}}\!\mbox{[} \, V_{0k}V_{k0}V_{0l}V_{l0} \, \mbox{]}\!\mbox{\textbf{]}}}{ \Delta_k^2 \Delta_l}  }  $ 
    \\
    &\\[-3.8mm]
    \hline
    & & & \\[-3.3mm] 
    &
    $ a^\dagger a a^\dagger b \, $ &  $  \omega_a - \omega_b  =  0 $ &
    $ \, {\displaystyle \hat{E}_{31}^{(0)}  = \sum_{\substack{l,k,q \not= 0 \\ q \not= k,l}}  \frac{\mathsf{ALLP}\mbox{\textbf{[}}\!\mbox{[} \, V_{0k}V_{kq}V_{ql}X_{l0} \,  \mbox{]}\!\mbox{\textbf{]}}}{\Delta_q \Delta_k \Delta_l} - \sum_{k,l\not= 0}  \frac{\mathsf{ALLP}\mbox{\textbf{[}}\!\mbox{[} \, V_{0k}V_{k0}V_{0l}X_{l0} \, \mbox{]}\!\mbox{\textbf{]}}}{ \Delta_k^2 \Delta_l} }  $ 
    \\
    &\\[-3.8mm]
    \hline
    & & & \\[-3.3mm] 
    Cross-Kerr $\,$ &
    $ a^\dagger a b^\dagger b \, $ &  \mbox{none} & 
    $ \, {\displaystyle \hat{E}_{22}^{(0)}  = \sum_{\substack{l,k,q \not= 0 \\ q \not= k,l}}  \frac{\mathsf{ALLP}\mbox{\textbf{[}}\!\mbox{[} \, V_{0k}V_{kq}X_{ql}X_{l0} \,  \mbox{]}\!\mbox{\textbf{]}}}{\Delta_q \Delta_k \Delta_l} - \sum_{k,l\not= 0}  \frac{\mathsf{ALLP}\mbox{\textbf{[}}\!\mbox{[} \, V_{0k}V_{k0}X_{0l}X_{l0} \, \mbox{]}\!\mbox{\textbf{]}}}{ \Delta_k^2 \Delta_l} }  $ 
    \\
    &\\[-3.8mm]
    \hline
    & & & \\[-3.3mm] 
    &
    $ a^\dagger a b^\dagger c \, $ &  $\omega_b - \omega_c  =  0 $ &
    $ \, {\displaystyle \hat{E}_{211}^{(0)}  = \sum_{\substack{l,k,q \not= 0 \\ q \not= k,l}}  \frac{\mathsf{ALLP}\mbox{\textbf{[}}\!\mbox{[} \, V_{0k}V_{kq}X_{ql}Y_{l0} \,  \mbox{]}\!\mbox{\textbf{]}}}{\Delta_q \Delta_k \Delta_l} - \sum_{k,l\not= 0}  \frac{\mathsf{ALLP}\mbox{\textbf{[}}\!\mbox{[} \, V_{0k}V_{k0}X_{0l}Y_{l0} \, \mbox{]}\!\mbox{\textbf{]}}}{ \Delta_k^2 \Delta_l} }  $ 
    \\
    &\\[-3.8mm]
    \hline
    & & & \\[-3.3mm] 
    Four-wave mixing $\,$ &
    $ a^\dagger b c^\dagger d \,$ & $\; \omega_a -  \omega_b + \omega_c - \omega_d =  0 \; $ &
    $ \, {\displaystyle \hat{E}_{1111}^{(0)}  = \sum_{\substack{l,k,q \not= 0 \\ q \not= k,l}}  \frac{\mathsf{ALLP}\mbox{\textbf{[}}\!\mbox{[} \, V_{0k}X_{kq}Y_{ql}Z_{l0} \,  \mbox{]}\!\mbox{\textbf{]}}}{\Delta_q \Delta_k \Delta_l} - \sum_{k,l\not= 0}  \frac{\mathsf{ALLP}\mbox{\textbf{[}}\!\mbox{[} \, V_{0k}X_{k0}Y_{0l}Z_{l0} \, \mbox{]}\!\mbox{\textbf{]}}}{ \Delta_k^2 \Delta_l} }  $     
    \\
    \hline
    \end{tabular} 
    \caption{Expressions for terms in the TIPT eigenvalue expansion, denoted $E^{(n)}_{jk\ldots}$, that determine the nonlinearities induced by coherently-driven multi-level emitters. The number of subscripts on the expansion term is the number of field modes in the corresponding nonlinear term, and the value of each subscript gives the number of mode operators for that mode contained in the nonlinear term. For example, the term $\hat{E}_{211}$ is the coefficient that gives nonlinear terms containing two mode operators for mode $a$, one for each of modes $b$ and $c$. Which nonlinear terms will be active is determined not only by the expansion term itself, but by which satisfy their respective resonance conditions for the frequencies of the fields (see Section~\ref{Sec2F}). The notation $\mathsf{ALLP}\mbox{\textbf{[}}\!\cdot\!\mbox{\textbf{]}}$ is defined in Section~\ref{sec2C}.} 
    \label{tab1}
\end{table*} 

\subsection{The effective nonlinear Hamiltonian for the field} 
\label{SecHeff} 

For the purposes of using an emitter to generate effective nonlinearities for up to four field modes we couple each mode to a different emitter transition, and all the couplings are perturbative from the point of view of the emitter. The Hamiltonian thus has the form given in Eqs.\ (\ref{H4p}), (\ref{Vform1}), and (\ref{Vform2}). At zero temperature (which for optical fields is equivalent to room temperature) the joint system will be in the subspace defined by the perturbed ground state eigenvector, $|0\rangle$. As explained above, the action of the Hamiltonian on this subspace is given by the field operator $\hat{E}^{(0)}$ so that the effective Hamiltonian for the field is 
\begin{align}
    H_{\ms{eff}} = \hat{E}^{(0)} = \sum_{j,k,l,q} (\lambda^j \nu^k \eta^l \xi^q) \hat{E}^{(0)}_{jklq}  . 
    \label{Heffem}
\end{align} 
Each term in the expansion on the RHS is a product of the field interaction operators. More specifically, a term $\hat{E}^{(0)}_{jklq}$ contributes a product of $m = j+k+l+q$ mode operators, and thus a specific nonlinearity. The first-order terms ($m=1$) vanish as described above. The second-order terms ($m=2$) give either linear interactions between the modes or frequency shifts of individual modes. The terms of third and higher order ($m \geq 3$) generate nonlinearities. Terms that are $m^{\msi{th}}$-order in the perturbation expansion correspond to $(m-1)^{\msi{th}}$-order  nonlinearities. Thus the expansion terms we give in Table~\ref{tab1} are sufficient to describe the generation of all nonlinearities up to third order with up to four fields. The reason that the order of the nonlinearity is one less than the order of the expansion is that the former is defined with reference to the differential equation for the electric field rather than the Hamiltonian (and we are now stuck with it).  

Examining the expressions for the expansion terms $E_{jklq}$ in Table~\ref{tab1} we see that the value of each of the subscripts ($j,k,l,q$) tells us how many times the elements of the corresponding interaction operator (one of $V$, $X$, $Y$, $Z$) appears in the products that make up that term. Since each element of $V$, $X$, $Y$, $Z$ contains, respectively, a mode operator for mode $a$, $b$, $c$, and $d$, the subscripts tell us how many mode operators for each of the modes appear in the nonlinearity generated by that term. They do not, however, tell us whether the mode operators are annihilation or creation operators or a mix of both. That depends on exactly which elements of the interaction operators contribute to the term and whether these are upper or lower diagonal (raising or lowering). In Table~\ref{tab1} we give examples of the nonlinear terms that can be generated by each expansion term. We have written these for the ground-state subspace, but those for the other subspaces are obtained merely by replacing $\hat{E}^{(0)}_{ijkl}$ with $\hat{E}^{(n)}_{ijkl}$. 

\subsection{Size of the coupling rates}
\label{Sec2Eb}

As we discussed in Section~\ref{sec2B}, perturbation theory is only valid so long as the power series in the perturbation parameters converges. For $j,k,l,q$ all greater than unity, each term in the power series expansion for the ``eigenvalues" (e.g., the expansion for $\hat{E}^{(0)}$ in Eq.(\ref{Heffem}) above) has the form 
\begin{align}
    & (\lambda^j \nu^k \eta^l \xi^q) \hat{E}^{(n)}_{jklq} 
    \sim \left( \frac{\lambda^j |V|^j}{\Delta^{j-1}} \right) \left( \frac{\nu^k |X|^k}{\Delta^{k-1}} \right) \left( \frac{\eta^l |Y|^l}{\Delta^{l-1}} \right) \left( \frac{\xi^q |Z|^q}{\Delta^{q-1}} \right) \nonumber \\
     & \;\;\;\;\; \sim \left( \frac{\lambda^j \langle a^\dagger a\rangle^{j/2}}{\Delta^{j-1}} \right) \left( \frac{\nu^k \langle b^\dagger b \rangle^{k/2}}{\Delta^{k-1}} \right) \left( \frac{\eta^l \langle c^\dagger c\rangle^{l/2}}{\Delta^{l-1}} \right) \left( \frac{\xi^q \langle d^\dagger d\rangle^{q/2}}{\Delta^{q-1}} \right) 
\end{align}
in which $\Delta$ is on the order of the detunings (or the transition frequencies if in the bare coupling regime) and $|V|$, $|X|$, \textit{etc}., are on the order of the typical matrix elements of the respective interaction operators, $V$, $X$, \textit{etc}. A sufficient condition for the convergence of the power series is that each of the products in the above expression are less than unity. This will be true if $\lambda$ satisfies 
\begin{align}
    \lambda \ll \frac{\Delta}{\sqrt{\langle a^\dagger a\rangle}} ,  \label{lambound}
\end{align} 
and similarly for the coupling rates $\nu$, $\eta$, $\xi$. These conditions on the coupling rates are sufficient for the perturbation series to converge but constitute bounds on the coupling rates only if they prove also to be \text{necessary} for this convergence. For single emitters this appears always to be the case, but for ensembles as we will see later the situation is more complex. 

\subsection{Selection of nonlinearities by resonance conditions and the rotating-wave approximation} 
\label{Sec2F} 

By choosing the right level structure and driving configuration one can tailor the coefficients of the power series for $\hat{E}^{(0)}$ to choose which nonlinearities will be generated. Below we apply our theory to the scheme of Schmidt and Imamo{\=g}lu in which a 4-level emitter is driven so as to generate a cross Kerr nonlinearity without the associated self-Kerr nonlinearities for each of the modes. 

Driving the emitter to tailor the interaction operators $V$, $X$, etc. is not the only mechanism that selects which nonlinearities will be active. Let us denote the frequency of mode $a_l$ by $\omega_l$. A nonlinear term of the form 
\begin{align}
    \mathcal{N} = a_1^{j_1} a_1^{\dagger k_1} a_2^{j_2} a_2^{\dagger k_2} \cdots a_n^{j_n} a_n^{\dagger k_n}   
\end{align}
has time dependence given by $e^{-i\Omega t}$ in the interaction picture where 
\begin{align}
    \Omega = (j_1-k_1) \omega_1 + (j_2-k_2) \omega_2 + \cdots + (j_n-k_n) \omega_n .
\end{align}
If $\Omega$ is much larger than the rate of the evolution generated by the nonlinear term itself, then the oscillation at frequency $\Omega$ will cause this effect to average to zero. Since the frequencies of optical modes are typically much larger than the rates of emitter-generated nonlinearities, the latter only survive under the condition that $\Omega = 0$. We note that for the terms $a^\dagger a$ (frequency shift), $(a^\dagger a)^2$ (self-Kerr), and $a^\dagger a b^\dagger b$ (cross-Kerr), the oscillation frequency $\Omega$ is always zero regardless of the  frequencies of the modes. Conversely, the oscillation frequency of $a^\dagger a^2$ can only be zero if the frequency of the field mode is zero, so this term is not typically active. 

\subsection{Full nonlinear dynamics of the field: the master equation}
\label{sec2G} 

We can now derive the effective Hamiltonian for the field, but we also need to include the effect of spontaneous decay of the emitter energy levels. For this we need to construct the effective master equation for the field. The spontaneous decay of the emitter is described by a Lindblad master equation for the emitter density matrix, $\rho_{\ms{e}}$~\cite{McCauley20b}: 
\begin{align}
    \dot\rho_{\ms{e}} & = \frac{-i}{\hbar}[\tilde{H},\rho_{\ms{e}} ] + \sum_{n} \frac{\gamma_{n}}{2} \left( 2 \tilde{\sigma}_{n0}\rho_{\ms{e}} \tilde{\sigma}^{\dagger}_{n0} - \tilde{\sigma}^{\dagger}_{n0} \tilde{\sigma}_{n0} \rho_{\ms{e}} - \rho_{\ms{e}} \tilde{\sigma}^{\dagger}_{n0} \tilde{\sigma}_{n0} \right) \nonumber 
    \label{fem} 
\end{align}
in which the transition operators are $\tilde{\sigma}_{n0} = \ket{\tilde{0}}\bra{\tilde{n}}$ and $\gamma_n$ is the decay rate from level $|\tilde{n}\rangle$ to the ground state. Here we restrict our analysis to levels that decay directly to the ground state. Decay to other levels has a more complex effect on the effective dynamics which we will discuss in Section~\ref{sec2H}. 

% The last two terms describe the reduction in the population of the levels that decay and the first term describes the re-population of the ground state. The last two terms can be absorbed into the Hamiltonian so that it becomes non-Hermitian: 
% \begin{align}
%     \dot\rho & = \frac{-i}{\hbar}\left( \tilde{H}_{\ms{nh}}\rho - \rho H_{\ms{nh}}^\dagger \right) + \sum_{n} \gamma_{n}  \tilde{\sigma}_{n0}\rho \tilde{\sigma}^{\dagger}_{n0}  , 
%     \label{fem2}
% \end{align}
% where the non-Hermitian Hamiltonian is 
% \begin{align}
%   H_{\ms{nh}} = H - i \sum_n \frac{\gamma_n}{2} \tilde{\sigma}^{\dagger}_{n0} \tilde{\sigma}_{n0} = H - i \sum_n \frac{\gamma_n}{2} \ket{\tilde{n}} \bra{\tilde{n}} . 
% \end{align}
% We can use the non-Hermitian Hamiltonian when we transform to the eigenbasis $\{|n\rangle\}$ that includes the classical driving, in which case the eigenstates $|n\rangle$ are unchanged but the eigenvalues $E_n$ become complex. We can use these complex eigenvalues when performing the perturbation expansion if we wish. In this case the resulting effective Hamiltonian for the field includes the damping in its own eigenvalues. Alternatively we can use the real (Hermitian) Hamiltonian for the perturbation expansion, in which case the effective Hamiltonian is Hermitian and the rest of the master equation can be constructed separately. 

As discussed above, so long as the emitter and the field remain in the subspace defined by the perturbation expansion for emitter state $|n\rangle$ (a subspace that contains essentially all field states) the effective dynamics for the field is given by the field operator generated by the expansion for the eigenvalue of $|n\rangle$. From the point of view of practicality, since the field and emitter can be assumed to be in the ground perturbed subspace at zero temperature, it is best to take this subspace, represented by the perturbed emitter state $|0\rangle$, as the one that the emitter and field will (approximately) remain in. To construct the master equation we need to determine the action of the transition operators $\sigma_{n0}$ on the subspace $|0\rangle$. To do this we first transform these operators to the driven basis and then project them onto this subspace. The effective transition operators are thus 
\begin{align}
    \sigma_{n}^{\ms{eff}} = \ket{0} \bra{0} U \ket{\tilde{0}} \bra{\tilde{n}} U^\dagger  \ket{0} \bra{0} , 
\end{align}
where $\ket{0} \bra{0}$ denotes the projector onto the subspace $|0\rangle$ (further details are given in Appendix~\ref{appA}). Since the effective dynamics remains in the subspace $\ket{0}$, we obtain the action of the transition operators in this subspace by discarding the outer projectors $\ket{0}$ and $\bra{0}$ in the expression above for $\sigma_{n}^{\ms{eff}}$. This gives us the effective transition operators purely as operators that act only on the field. Denoting these effective transition operators by $\Sigma_n$ we have   
\begin{align} 
    \Sigma_{n} =  \bra{0} U \ket{\tilde{0}} \bra{\tilde{n}} U^\dagger  \ket{0}  . 
    \label{sigman}
\end{align}

Under the assumption that the driving does not couple the bare ground state of the emitter, $\ket{\tilde{0}}$, to the other levels we have $U \ket{\tilde{0}} = \ket{0_0}$ (note that the latter is defined in Eq.(\ref{nlambda}) and to second order the inner product $\bra{0} 0_0 \rangle$ for one field is 
\begin{align}
   \bra{0} \! 0_0 \rangle & = 1 + \lambda^2 \langle 0_{2} | 0_0 \rangle + \ldots  
\end{align}
and for two fields it can be written as 
\begin{align}
    \bra{0}\! 0_0 \rangle  & = 1 + 
    \left( \begin{array}{c} \lambda \\ \nu \end{array} \right)^{\ms{t}}
    \left( \begin{array}{cc}
     \langle 0_{02} | 0_0 \rangle    & \langle 0_{11} | 0_0  \rangle/2  \\
     \langle  0_{11} | 0_0 \rangle/2    & \langle 0_{20} | 0_0  \rangle 
    \end{array} \right) 
    \left( \begin{array}{c} \lambda \\ \nu \end{array} \right) + \ldots  
\end{align} 
In this case the inner product $\bra{\tilde{n}} U^\dagger  \ket{0} $ has no zeroth-order component, and so for two fields is   
\begin{align}
    \bra{\tilde{n}} U^\dagger  \ket{0} & = \sum_{j=1}^{\infty} \sum_{k=1}^{\infty} \lambda^j \nu^k  \bra{\tilde{n}} U^\dagger  \ket{0_{jk}} . 
\end{align}
Since the leading-order term in the transition operators is first order in the field operators, and all terms in the master equation contain a product of two transition operators, to construct the master equation up to $4^{\msi{th}}$ order we need ``only" calculate the transition operators to $3^{\msi{rd}}$ order, which in turn means that we only need to determine $\bra{0} U \ket{0}$ to $2^{\msi{nd}}$ order. Once we have obtained the effective transition operators, $\Sigma_n$, to the desired order, the effective master equation for the field density matrix, $\rho$, is 
\begin{align}
    \dot\rho & = \frac{-i}{\hbar}[H_{\ms{eff}}^{(0)},\rho] + \sum_{n} \frac{\gamma_{n}}{2} \left( 2 \Sigma_n\rho \Sigma^{\dagger}_n - \Sigma^{\dagger}_n \Sigma_n \rho - \rho \Sigma^{\dagger}_n \Sigma_n \right) 
    \label{effme} 
\end{align}
where $H_{\ms{eff}}$ is the effective Hamiltonian generated by the emitter, Eq.(\ref{Heffem}).

\subsection{Preparing the appropriate initial state}
\label{sec2H}

We saw in Section~\ref{SecHeff} that the effective Hamiltonian for the field is given by $\hat{E}^{(0)}$ so long as the joint system is in a state that lies in the subspace denoted by the perturbed eigenvector  $\ket{0}$. If we assume that the emitter and the field begin in their joint ground state, then since this state is (the lowest energy state) in the subspace $\ket{0}$, the effective Hamiltonian for the field is $\hat{E}^{(0)}$. However, this nonlinear evolution is hardly useful unless we can prepare the field in other initial states. Certainly the subspace $\ket{0}$ contains (almost) all field states, so in theory we can prepare any initial field state while remaining in this space. One way to do this is to apply a time-dependent Hamiltonian that changes the state of the field slowly compared to the energy gap between the populated part of the subspace $\ket{0}$ and the subspace $\ket{1}$. In this case the joint system will remain in the subspace $\ket{0}$ by the adiabatic theorem. We now show that the use of this adiabatic evolution, or some even more sophisticated procedure, is in fact necessary; if one applies a field Hamiltonian that transforms the initial ground state to some other state too quickly, then the resulting joint state has a component outside the subspace $\ket{0}$. 

The situation described by the Hamiltonian $H_{\ms{B}}$ in Section~\ref{sec2A}, in which the emitter/field interaction is a product of an operator of the emitter and a field mode, allows us to obtain considerable insight into the joint eigenstates. We can write $H_{\ms{B}}$ (Eq.(\ref{HA})) as 
\begin{align}
    \mathcal{H}  = H_0 +  G \Lambda + H_{\ms{f}} 
\end{align} 
with $G$ and $\Lambda$ are operators of the emitter and the field mode, respectively. We define the eigenvalues and eigenstates of the mode operator $\Lambda$ by $\Lambda \ket{\lambda} = \lambda \ket{\lambda}$. For each subspace defined by the mode state $|\lambda\rangle$ the Hamiltonian of the emitter is $\mathcal{H}_\lambda  = H_0 + \lambda G$ and we define the eigenvectors of $\mathcal{H}_\lambda$ as 
\begin{align} 
      \mathcal{H}_\lambda \ket{n^{(\lambda)}}  = \hat{E}^{(n)}(\lambda) \ket{n^{(\lambda)}} . 
\end{align}
The eigenvectors of $H_0 + G \Lambda$ are now given by $\ket{n^{(\lambda)}} \ket{\lambda}$. In particular, we have  
\begin{align}
 (H_0 + G \Lambda) \ket{n^{(\lambda)}} \ket{\lambda} & = \hat{E}^{(n)}(\lambda) \ket{n^{(\lambda)}}  \ket{\lambda} \nonumber \\
   & = \Biggl( \sum_j \hat{E}^{(n)}_j \lambda^j \Biggr)  \ket{n^{(\lambda)}}  \ket{\lambda} \\
   & = \hat{E}^{(n)}(\Lambda) \ket{n^{(\lambda)}}  \ket{\lambda} . 
\end{align} 
Here $\sum_j \hat{E}^{(n)}_j \lambda^j$ is the perturbation expansion for the emitter eigenvalues when the perturbation is $\lambda G$. Now consider the effective Hamiltonian for the field when,  for each eigenstate of $\Lambda$, the emitter is in its corresponding eigenstate $\ket{n^{(\lambda)}}$:  
\begin{align}
    H_{\ms{f}}^{(n)} = \hat{E}^{(n)}(\Lambda) + H_{\ms{f}} .  
\end{align}
If we write the eigenstates of this Hamiltonian as  
 \begin{align}
    |f_m^{(n)}\rangle = \sum_\lambda c_{m\lambda}^{(n)}  \ket{\lambda} , \;\;\; m = 0,1,2, \ldots 
\end{align}
then we see that the states 
\begin{align}
    |J_m^{(n)}\rangle = \sum_\lambda c_{m\lambda}^{(n)}  \ket{n^{(\lambda)}}  \ket{\lambda}
    \label{Jstates}
\end{align}
are the eigenstates of the joint system. Having constructed these eigenstates, we can write the subspace $\ket{0}$ more explicitly as the space spanned by the set of states 
\begin{align}
    \Bigl\{|J_m^{(0)}\rangle : m = 0,1,2,\ldots \Bigr\}. 
\end{align} 

Having established the above form for the joint eigenstates, we can examine what happens if we change the state of the field suddenly. A simple but instructive example is that of displacing the eigenvalues of $\Lambda$. If we start in the joint ground state and displace the eigenstates by an amount $x$ the new joint state is 
\begin{align}
    |J_0^{(0)}\rangle_x = \sum_\lambda c^{(0)}_{0\lambda}  \ket{0^{(\lambda)}} \ket{\lambda + x} . 
\end{align} 
This new state no longer lies fully within the emitter ground-state subspace. To determine how much of the higher emitter states are mixed in we need the overlaps 
\begin{align}
   \langle n^{(\lambda+x)} \ket{0^{(\lambda)}} & = \Biggl(\sum_j \bra{n_j} (\lambda + x)^j\Biggr) \Biggl(\sum_k  \lambda^k \ket{0_k} \Biggr) \nonumber \\
   & = (\lambda + x) \langle n \ket{0_1} + \lambda \langle n_1 \ket{0} + \mathcal{O}(\lambda^2) \nonumber \\
   & = (\lambda + x) \frac{V_{n0}}{\Delta_{n0}} - \lambda \frac{V_{n0}}{\Delta_{n0}} + \mathcal{O}(\lambda^2)  \\
   & = x \frac{V_{n0}}{\Delta_{n0}} + \mathcal{O}(\lambda^2)
\end{align}
where we have used the fact that $V_{0n} = V_{n0}^*$ and $\Delta_{0n} = -\Delta_{n0}$. Thus when the field is changed rapidly there are components of the joint state that contain emitter excited states, where the size of these components is first-order in the perturbation and proportional to the change in the state of the field. The components that contain the emitter excited-state subspaces will induce nonlinear evolution generated by $\hat{E}^{(n)}$ for $n>0$, which may be different to that generated by the ground state. 

\begin{table}[t]
\textbf{Procedure to calculate nonlinearities for a single emitter} 
\vspace{1mm}
 \hrule
    \begin{enumerate}
    \item Determine the emitter matrix elements of the interaction operators. These matrix elements are proportional to annihilation and creation operators for the field modes
    \item Determine the emitter dressed states 
    \item Transform the matrices for the interaction operators to this dressed-state basis 
    \item Substitute the elements of these transformed matrices into the formulae in Table~\ref{tab1} 
    %\\
    %4 & If desired, use Eq.(\ref{sigman}) to calculate the lindblad operators 
\end{enumerate}
 \hrule
\caption{The dressed states referred to here are the eigenstates of the emitter Hamiltonian including the classical driving fields. An example of applying the above procedure is given in Section~\ref{secSI}.\label{tab2}}
\end{table}

\section{Quantum Treatment of the Schmidt-Imamo{\=g}lu cross-Kerr scheme}
\label{secSI}

We give the steps involved in calculating the nonlinearities generated by a single emitter in Table~\ref{tab2}. As an example of using this method we treat the Schmidt-Imamo{\=g}lu EIT scheme that generates the cross-Kerr nonlinearity~\cite{Schmidt96}. This nonlinearity could potentially be used to realize number-resolving QND measurements of photons. In this measurement scheme, first suggested by Imoto, Haus, and  Yamamoto~\cite{Imoto85}, a mode containing a sufficiently large coherent state interacts via the cross-Kerr effect with a mode containing only a few photons. The cross-Kerr effect generates a photon-number-dependent phase shift of the coherent state, which can then be measured with homodyne detection. For the purpose of the measurement, so long as the coherent state is shot-noise limited, its amplitude can be increased to offset the weakness of the cross-Kerr nonlinearity. (In the Supplement we  give the simplest example, calculating the nonlinearities generated by an undriven two-level system.)

\begin{figure}[t]
\centering
\leavevmode\includegraphics[width = 0.7 \columnwidth]{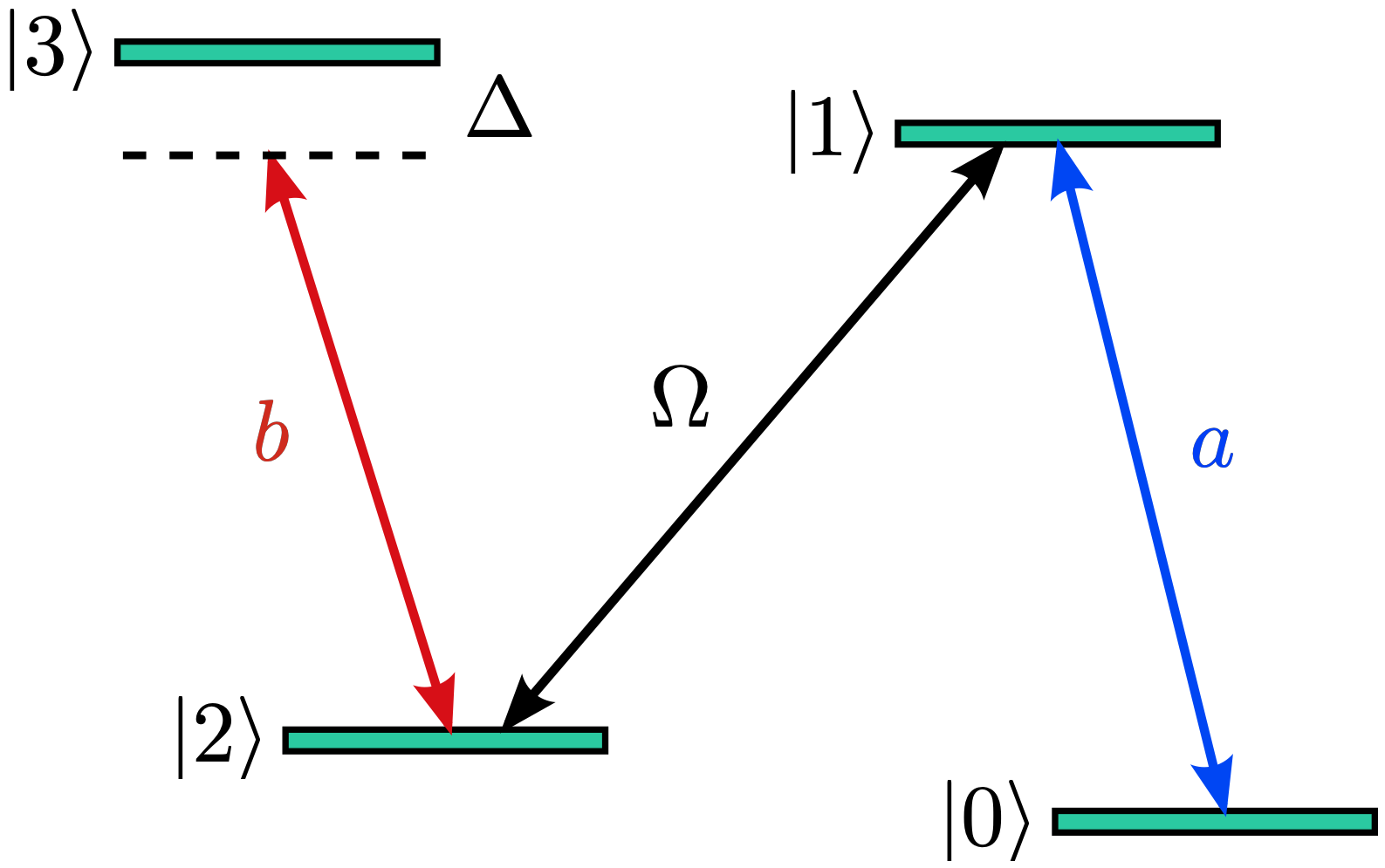}
\caption{(Color online) The level structure and driving scheme of the Schmidt-Imamo{\=g}lu method for producing a cross-Kerr nonlinearity without the associated self-Kerr nonlinearities. The states $\ket{0}$ and $\ket{2}$ are ground states, while $\ket{1}$ and $\ket{3}$ are excited states with significant decay rates. The scheme benefits from the EIT mechanism in which the effect of the decay rate of level $\ket{1}$ is essentially eliminated. Here mode $a$ couples to the transition $\ket{0}\leftrightarrow \ket{1}$ (on resonance) and $b$ to $\ket{2}\leftrightarrow \ket{3}$ (with detuning $\Delta$). The $\ket{2}\leftrightarrow \ket{1}$ transition is classically driven with Rabi frequency $\Omega$; it is this drive that implements EIT, mixing levels $\ket{1}$ and $\ket{2}$.} 
\label{fig:scheme} 
\end{figure}

The key problem with the use of bulk crystal nonlinearities for this QND measurement scheme is that the self-Kerr nonlinearity, which is always active in a material with $3^{\msi{rd}}$-order nonlinearities, swamps the phase-shift signal due to the cross-Kerr nonlinearity~\cite{Balybin22}. What is needed therefore is a way to create a cross-Kerr nonlinearity, given by $a^\dagger a b^\dagger b$, in which at least one of the modes has no self-Kerr. A particularly effective way to do this was devised by Schmidt and Imamo{\=g}lu~\cite{Schmidt96}. Their scheme, which we depict in Fig.\ref{fig:scheme}, uses a four-level emitter driven in a way that takes advantage of EIT, in which the spontaneous decay from one of the levels is effectively eliminated through destructive interference. 

In the Schmidt-Imamo{\=g}lu scheme the emitter is coupled to two quantum-mechanical field modes so the Hamiltonian in the interaction picture, after applying the rotating-wave approximation, has the form  
\begin{align}
    H & = \tilde{H}_0 +  \lambda \tilde{V} +  \nu \tilde{X}  
\end{align} 
with 
\begin{align}
   \tilde{V} & =  \hbar \sigma_{10}^\dagger  a + \hbar \sigma_{10} a^\dagger  , \\
   \tilde{X} & =  \hbar \sigma_{32}^\dagger  b + \hbar \sigma_{32} b^\dagger   , 
\end{align}
and where $a$ and $b$ are the mode operators for the respective field modes. Here $\lambda$ and $\nu$ are the coupling rates between the respective cavity modes and the emitter, and so $V$ and $X$ have dimensions of angular momentum. The Hamiltonian of the driven emitter and the interaction operators, in the interaction picture, are 
\begin{align} 
    \frac{\tilde{H}_0}{\hbar} = \left( \begin{array}{cccc}
               0 &       0 &            0 &       0   \\
               0 &       0 &       \Omega &       0   \\
               0 &  \Omega &            0 &       0   \\ 
               0 &       0 &            0 &   \Delta 
               \end{array}  \right) ,     
    \frac{\tilde{V}}{\hbar} = \left( \begin{array}{cccc}
               0 &       a^\dagger &       0 &       0   \\
               a &       0 &       0 &       0   \\
               0 &       0 &       0 &       0   \\ 
               0 &       0 &       0 &       0  
               \end{array}  \right) , 
    \frac{\tilde{X}}{\hbar} = \left( \begin{array}{cccc}
               0 &       0 &       0 &       0   \\
               0 &       0 &       0 &       0   \\
               0 &       0 &       0 &      b^\dagger   \\ 
               0 &       0 &       b &       0  
               \end{array}  \right)  \nonumber 
\end{align}
Here $\Delta$ is the detuning of the field mode $b$ from the upper transition. We first transform to the basis that diagonalizes $H_0$, for which the transformation is  
\begin{align}
    U       =  \left( \begin{array}{cccc}
               1  &       0 &         0 &       0   \\
               0 &  -1/\sqrt{2}  & 1/\sqrt{2}  &       0   \\
               0 &  1/\sqrt{2}  & 1/\sqrt{2}  &       0   \\ 
               0 &       0 &         0 & 1  
               \end{array}  \right) 
\end{align}
The transformed Hamiltonian and interaction operators are  
\begin{align}
          H_0 = \hbar \left( \begin{array}{cccc}
               0 &       0 &       0 &       0   \\
               0 & \! -\Omega \!\!  &       0 &       0   \\
               0 &       0 & \! \Omega \! &       0   \\ 
               0 &       0 &       0 &  \Delta  \! 
               \end{array}  \right) , 
\end{align}
with 
\begin{align}
          V =  \frac{\hbar}{\sqrt{2}}
               \left( \begin{array}{cccc}
               0 &   -a^\dagger \! \! &    a^\dagger &       0   \\
               -a \!\! &       0 &       0 &       0   \\
               a &       0 &       0 &       0   \\ 
               0 &       0 &       0 &       0  
               \end{array}  \right)  , \;\;\; 
          X =  \frac{\hbar}{\sqrt{2}}
               \left( \begin{array}{cccc}
               0 &       0 &       0 &       0   \\
               0 &       0 &       0 &   \! b^\dagger    \\
               0 &       0 &       0 &   \!  b^\dagger   \\ 
               0 &  \!    b \! &    \!  b  &       0  
               \end{array}  \right) . 
              \nonumber
\end{align}

We can now calculate the sizes of the cross-Kerr and two self-Kerr nonlinearities using the expressions for $\hat{E}^{(0)}_{22}$ and $\hat{E}^{(0)}_{40} \equiv \hat{E}^{(0)}_{4}$ in Table~\ref{tab1}. (The expression for $\hat{E}^{(0)}_{04}$ is merely $\hat{E}^{(0)}_{40}$ with $V$ replaced by $X$.) Examining the expression for $\hat{E}^{(0)}_{22}$ we find that almost all the terms vanish because $X$ has no elements that couple to state $|0\rangle$. We have  
\begin{align} 
    \hat{E}^{(0)}_{22} 
    & = \sum_{\substack{l,q,p \not= 0 \\ q \not= l,p}} \frac{  V_{0l} X_{lq}  X_{qp} V_{p0} }{ \Delta_{l} \Delta_{q} \Delta_{p}} =   \frac{  V_{01} X_{13} X_{31} V_{10} }{ \Delta_{1} \Delta_{3} \Delta_{1}} \nonumber \\
    & \;\;\; + \frac{  V_{01} X_{13} X_{32} V_{20} }{ \Delta_{1} \Delta_{3} \Delta_{2}} + \frac{  V_{02} X_{23} X_{31} V_{10} }{ \Delta_{2} \Delta_{3} \Delta_{1}} + \frac{  V_{02} X_{23} X_{32} V_{20} }{ \Delta_{2} \Delta_{3} \Delta_{2}} \nonumber \\
    & =  - \frac{\hbar }{\Delta\Omega^2} a^\dagger a b^\dagger b  . 
\end{align}
For $\hat{E}^{(0)}_{40}$ the first term vanishes and we have 
\begin{align}
    \hat{E}^{(0)}_{40} & =  - \left(\sum_{l\not= 0}  \frac{V_{0l}V_{l0} }{ \Delta_{l}}  \right)  \left(\sum_{l\not= 0}  \frac{V_{0l}V_{l0} }{ \Delta_{l}^2}  \right)  \nonumber \\
   & = \frac{\hbar}{4}\left( \frac{1}{-\Omega} +  \frac{1 }{\Omega} \right)  \left(  \frac{1}{\Omega^2} + \frac{1}{\Omega^2} \right) \left( a^\dagger a\right)^2
   = 0  . 
\end{align}
The self-phase modulation for the mode $b$ is automatically zero because $X$ has no elements that couple $|0\rangle$ to the other states. 

The Schmidt-Imamo{\=g}lu scheme generates no frequency shifts or second-order nonlinearities because the second- and third-order expansion coefficients are all zero. The effective Hamiltonian to fourth order in the perturbation parameters $\lambda/\Omega$, $\lambda/\Delta$, $\nu/\Omega$, and $\nu/\Delta$, is therefore
\begin{align}
   H_{\ms{SI}} & = \hbar \omega_a a^\dagger a +  \hbar\omega_b  b^\dagger b  +  \lambda^2 \nu^2 \hat{E}^{(0)}_{22}   \nonumber \\ 
    & = \hbar \omega_a a^\dagger a + \hbar \omega_b  b^\dagger b  - \hbar\kappa a^\dagger a b^\dagger b
\end{align} 
where the rate of the cross-Kerr nonlinearity is 
\begin{align}
    \kappa = \lambda \left(\frac{ \lambda}{\Delta}\right) \left(\frac{\nu}{\Omega} \right)^2 . 
    \label{crossKrate}
\end{align}
To complete the description of the effective dynamics, given by the master equation in Eq.(\ref{effme}), we need to calculate the effective transition operators using Eq.(\ref{sigman}): 
\begin{align} 
    \Sigma_2 & = \bra{0} \! 0_0\rangle \langle \tilde{2}  \! \ket{0}  , \\
    \Sigma_3 & = \bra{0} \! 0_0\rangle \langle \tilde{3}  \!   \ket{0} .  
\end{align} 
These transition operators have the respective damping rates $\gamma_2$ and $\gamma_3$. Calculating the exact eigenvalues for the emitter matrix $H + V$ shows that the ground state contains no component of $|\tilde{2}\rangle$ so  $\langle \tilde{2}\!\ket{0} = 0$. As a result $\Sigma_2 = 0$ so the effective dynamics is unaffected by the decay of level $|\tilde{2}\rangle$. This is precisely the EIT effect, induced by the driving that couples levels $|\tilde{1}\rangle$ and $|\tilde{2}\rangle$,  eliminating absorbsion by the latter. On the other hand, to determine $\langle \tilde{3}\! \ket{0} = \langle 3_0\! \ket{0}$ we have to calculate the coefficients $\langle 3_0\! \ket{0_{jk}}$ in the expansion 
\begin{align}
    \langle 3_0\! \ket{0}  = \sum_{j,k} \langle 3_0\! \ket{0_{jk}} \lambda^j \nu^k    . 
\end{align}
We find that up to $4^{\msi{th}}$ order the only non-zero coefficient is $\langle 3_0\! \ket{0_{11}} = (a^\dagger b + b^\dagger a)/(\Delta\Omega)$, and thus  
\begin{align} 
     \langle \tilde{3}\! \ket{0}  & = \frac{\lambda\nu}{\Delta\Omega} (a^\dagger b + b^\dagger a ) . 
\end{align} 
Since we only wish to calculate the effective master equation to $4^{\msi{th}}$ order, and $\langle \tilde{3}\! \ket{0}$ is already $2^{\msi{nd}}$ order, we need $\bra{0}\! 0_0\rangle$ only to zeroth order. The result is that
\begin{align}
    \Sigma_3 & = \frac{\lambda\nu}{\Delta\Omega} (a^\dagger b + b^\dagger a )  
\end{align}
Substituting this into the master equation, Eq.(\ref{effme}), and making the rotating-wave approximation (which eliminates all terms in which the number of creation and annihilation operators are different) we obtain 
\begin{align}
    \dot\rho & = \frac{-i}{\hbar}[H_{\ms{SI}},\rho] -  \frac{\gamma}{2} \left( 2 \Sigma_3\rho \Sigma^{\dagger}_3 - \Sigma^{\dagger}_3 \Sigma_3 \rho - \rho \Sigma^{\dagger}_3 \Sigma_3 \right) 
    \label{effme2} 
\end{align}
with
\begin{align}
    \gamma & = \gamma_3 \left( \frac{\lambda}{\Delta} \right)^2 \left( \frac{\nu}{\Omega} \right)^2 . 
\end{align}

\section{Generation of nonlinearities by ensembles} 
\label{secEns}

Up to this point all our analysis has been concerned with the generation of nonlinearities by a single emitter. We now consider applying the tools we have developed to an ensemble of $N$ identical independent emitters. This more complex scenario exhibits new phenomena. In particular the behavior of the two kinds of emitter/field interactions described in Section~\ref{sec2A}, those given by $H_{\ms{B}}$ (Eq.(\ref{HA})) and $H_{\ms{R}}$ (Eq.(\ref{baseH2})), is very different for large $N$. The first of these interactions is able to generate nonlinearities that scale with the number of emitters at the few-photon level, whereas the second is not. The fact that $H_{\ms{B}}$ is able to generate single-photon giant nonlinearities also reveals how EIT schemes in the regime of $H_{\ms{R}}$ are able to generate nonlinearities that scale with $N$ in the semiclassical limit.  

To understand how nonlinearities are generated by an ensemble of identical systems all interacting in the same way with a field mode, insight is provided by the simplest example, an ensemble of two-level emitters. We begin by considering this example, for the case in which the emitter/field interaction has the rotating-wave form.  

\subsection{An ensemble of two-level emitters in the \\ rotating-wave regime}
\label{tlsym} 

In the rotating wave regime, for which it is natural to use the interaction picture, the Hamiltonian for $N$ two-level systems (from now on, ``qubits") coupled to a field mode is 
\begin{align}
    H_{\ms{Q}} = \frac{\hbar\Delta}{2} \sum_{j=1}^N \sigma_z^{(j)} + \hbar \lambda \sum_{j=1}^N \bigl( \sigma^{(j)\dagger} a + \sigma^{(j)} a^\dagger  \bigr) , 
\end{align} 
where as usual $a$ is the mode annihilation operator. Given the fact that the field interacts with each emitter in an identical way, if we start with all the emitters in the ground state the field can only excite completely symmetric superpositions of the emitter states (joint states that are symmetric under any permutation of the emitters). The structure of this symmetric subspace has already been worked out as part of the theory of angular momentum: each qubit corresponds to a spin one-half system by virtue of having two levels; the symmetric space is the $(N+1)$-dimensional space with total angular momentum $N/2$. The states in this space can be labelled by the number of qubits in the excited state. Denoting this number by $n$ the states are $|n\rangle$ with $n = 0, 1, \ldots, N$. The action of the operator 
\begin{align}
    J_+ = \sum_j \sigma^{(j)\dagger} 
\end{align}
on this space is that of a (nonlinear) raising operator:
\begin{align}
    J_+ \ket{n} & = \sqrt{(N-n)(n+1)} \ket{n+1} . 
\end{align}
As the number $N$ increases, the lowest lying energy levels become more and more linear because the spacing between energy levels becomes increasingly uniform. This is a reflection of the fact that the dynamics of a large ensemble is approximately linear when the excitation of the atoms is low. 

Writing the Hamiltonian explicitly in the symmetric subspace we have $H = H_0 + \lambda V$ with 
\begin{align}
    H_0 & = \hbar \Delta \sum_{n=0}^N n \ket{n}\!\bra{n} , \\ V   & = \hbar \,\bigl( J_+ a + J_- a^\dagger \bigr)  , 
\end{align}
where $J_- \equiv J_+^\dagger$. We can now apply the TIPT machinery we developed in Section~\ref{sec2} directly to this Hamiltonian to determine the nonlinearities generated by the ensemble of qubits. To distinguish the expansion coefficients for the ensemble from those for a single emitter, we will denote the former using the blackboard bold font ($\mathbb{E}$). To fourth order the expansion coefficients for the ensemble are  
\begin{align}
    \mathbb{E}^{0}_2 & = \sum_{l\not= 0} \frac{V_{0l}V_{l0}}{ \Delta_l} = \frac{ J_- \ket{1} \bra{1} J_+ \ket{0}}{-\Delta} = - \hbar \left(\frac{N}{\Delta}\right) a^\dagger a  \\
    \mathbb{E}^{0}_3 & = 0 \\
    \mathbb{E}^{0}_4 & = \sum_{\substack{l,k,q \not= 0 \\ q \not= k,l}}  \frac{ V_{0k}V_{kq}V_{ql}V_{l0} }{\Delta_q \Delta_k \Delta_l} - \sum_{k,l\not= 0}  \frac{V_{0k}V_{k0} V_{0l}V_{l0}}{ \Delta_k^2 \Delta_l} \nonumber  \\ 
    & = - \frac{ V_{01}V_{12}V_{21}V_{10} }{-2 \Delta^3} -   \frac{V_{01}V_{10} V_{01}V_{10}}{ -\Delta^3} \nonumber \\
%  & = \frac{- N(N-1) a^{\dagger 2} a^2 +  N^2  (a^{\dagger} a)^2 }{ \Delta^3 } \nonumber \\
  & = \hbar \frac{ N }{ \Delta^3 } (a^{\dagger} a)^2 + \hbar \frac{ N(N-1) }{ \Delta^3 }  a^{\dagger} a  
\end{align}
The essential result here is that even though the symmetric subspace becomes increasingly linear as $N$ increases, it is nevertheless non-linear \textit{enough} to generate a nonlinearity, $\mathbb{E}^{(0)}_4$, that \textit{increases} with $N$. While the dynamics of the ensemble may be essentially linear when interacting resonantly with other systems (for low excitations), for perturbative interactions it becomes effectively \textit{more nonlinear} as $N$ increases. 

The effective Hamiltonian for the field mode, in the Schr\"{o}dinger picture, is 
\begin{align}
    H_{\ms{eff}} & = \hbar \lambda \left\{ \left[ \omega -  N  \epsilon_\lambda +  N (N-1)  \epsilon_\lambda^3 \right]  a^\dagger a  + N  \epsilon_\lambda^3 (a^{\dagger} a)^2 \right\} ,
    \label{HeffTLS} 
\end{align} 
with $\epsilon_\lambda \equiv \lambda/\Delta$. The self-Kerr nonlinearity is the coefficient of $(a^\dagger a)^2$ (divided by $\hbar$), and is thus 
\begin{align}
    \kappa_{\ms{s}} & = \lambda N  \epsilon_\lambda^3 = \frac{ N \lambda^4 }{ \Delta^3 } .
    \label{sKerr}
\end{align} 
We see that the size of the Kerr nonlinearity increases with $N$, and while the frequency shift also increases with $N$, there is an additional contribution to this shift at next-to-leading order in $\epsilon_\lambda$ that scales as $N^2$. 

Now recall that the interaction operator between the field and the emitters in the symmetric subspace is given by $J^{+}$, and that the resulting matrix element that couples the ground state to the first symmetric excited state contains a factor of $\sqrt{N}$. This is the effect discovered by HBP~\cite{Hartmann06,Hartmann07}, and given the bound in Eq.(\ref{lambound}) it strongly suggests that to remain in the perturbative regime for the ensemble of emitters requires that 
\begin{align}
    \lambda \ll \frac{\Delta}{\sqrt{N \langle a^\dagger a \rangle}} .  
    \label{limrN}
\end{align}
This would effectively eliminate the linear increase in the nonlinearity in Eq.(\ref{HeffTLS}) because $\lambda$ must be reduced as $N$ increases. We will show in Section~\ref{barecoup} that, remarkably, the $\sqrt{N}$ factor in the ground-state coupling for ensembles does not necessarily result in the bound given by Eq.(\ref{limrN}). %(It does, however, impose this limit for all emitter/field systems in the RWA regime, except when the field states are semi-classical.) 
First, however, we show how to calculate nonlinearities for ensembles of any multi-level emitter. 

% ***** The terms generated by the perturbative interaction with the emitter scale as $N$, which is the behavior assumed in the semiclassical theory. However, note that the matrix element of the interaction operator, $V$, that couples the ground state, $|0\rangle$, to the first symmetric excited state, $\ket{1}$, is equal to $\hbar \lambda \sqrt{N}$. This coupling is $\sqrt{N}$ larger than that for a single emitter. The perturbative expansion is only valid so long as the coupling between the eigenstates of $H_0$ is much less than the energy gap between adjacent states, meaning that 
% \begin{align}
%     \lambda \ll \frac{\Delta}{\sqrt{N}} .  
% \end{align}

% And if this were true, it would imply that ensembles are not able to generate the same size nonlinearities at the few-photon level as they do in the semiclassical limit in which the coherent states involved have many photons. 

% If the behavior at the few-photon level is not the same as that in the semiclassical limit, then this limiting behavior must emerge as the coherent amplitude of the field mode(s) is increased. However, examining how this behavior should emerge reveals a contradiction. The solution is, remarkably, that the behaviour is the same for a single photon as it is in the classical limit: the perturbative treatment of the ground state for the ensemble is, in fact, valid beyond the breakdown of perturbation theory applied to the symmetric subspace. This can be shown by applying perturbation theory separately to each of the ensemble members. 

\subsection{Applying TIPT to the symmetric subspace \\ of an ensemble of arbitrary emitters} 
\label{Arbsym}

We now show how to perform the above analysis for an ensemble of emitters with an arbitrary number of levels. To do so, one must calculate the matrix elements of the perturbative interaction in the symmetric subspace of the ensemble. By inspection we see that $n^{\msi{th}}$ order terms in the perturbation expansion for a single emitter involve excited states that can be reached from the ground state by taking $n/2$ steps, in which each step is a matrix element coupling two states. Translating this to an ensemble, $n^{\msi{th}}$-order terms in the perturbation expansion involve only matrix elements that connect the ground state to states in which at most $n/2$ emitters are in excited states. To calculate the perturbation expansion to $4^{\msi{th}}$ order we therefore need only the matrix elements of the interaction Hamiltonian in the subspace spanned by the ground state and symmetric states in which there are only one or two emitters in an excited state. 

Consider an ensemble of $N$ emitters, each with $M$ levels. As before we will denote the levels of a single emitter by $|j\rangle$, $j = 0,1, \ldots, M-1$, with $|0\rangle$ being the ground state. We will denote the state of the whole ensemble in which every emitter is in its ground state by $|\mbox{\textbf{0}}\rangle$. Let us also denote the symmetric states in which $n_j$ atoms are in excited state $\ket{j}$ by $|n_j\rangle$, and those in which $n_j$ atoms are in state $\ket{j}$ and $n_k$ states are in state $\ket{k}$ by $|n_j,n_k\rangle$. The interaction Hamiltonian between emitter $n$ and the field is $V_n = V$, so that the total interaction Hamiltonian is 
\begin{align}
    \mathcal{V} = \sum_{n=0}^{N-1} V_n  
\end{align}
where the matrix elements of each $V_n$ are $V_n^{(jk)} = V_{jk}$. Because this interaction Hamiltonian is symmetric under any permutation of the emitters, starting from the ground state it can only generate symmetric states. 

By applying an arbitrary $\mathcal{V}$ to the ground state a few times one can readily calculate the following matrix elements of $\mathcal{V}$: 
\begin{align}
  \bra{1_j} \mathcal{V} \ket{\mbox{\textbf{0}}} & = \sqrt{N} V_{j0} \label{rNVj0} \\
  \bra{(n+1)_j} \mathcal{V} \ket{n_j} & =  \sqrt{(N-n)(n+1)} V_{j0} \\
  \bra{1_k} \mathcal{V} \ket{1_j} & = V_{kj} \label{eqvjk} \\ 
    \bra{1_k, 1_j} \mathcal{V} \ket{1_j} & = \sqrt{N-1}  V_{k0} \\
  \bra{1_k,1_j}  \mathcal{V} \ket{2_j}  & =  \sqrt{2} V_{kj} \label{r2Vkj} \\
  \bra{1_l,1_j}  \mathcal{V} \ket{1_k,1_j}  & = V_{lk} \label{1l1j} . 
\end{align}
The above matrix elements include all those between symmetric states with no more than two atoms in an excited state. From Eq.(\ref{rNVj0}) we see that the coupling between the ground state and any level to which it is coupled by the single-emitter interaction, $V$, is increased by a factor of $\sqrt{N}$. This is the effect noted by HBP~\cite{Hartmann06, Hartmann07}, and it impacts the scaling of the effective nonlinearities in the RWA regime as we will show below. Note, however, that the coupling between any two levels that are not the ground state remain independent of $N$, and are at most modified only by a small factor (e.g. Eqs.(\ref{eqvjk}), (\ref{r2Vkj}), and (\ref{1l1j})), 

The matrix elements given in Eqs.(\ref{rNVj0}) through (\ref{1l1j}) are sufficient to calculate the nonlinearities generated by any ensemble up to $4^{\msi{th}}$ order. As an example, consider the self-Kerr nonlinearity generated for a single mode by an ensemble of arbitrary emitters. Using the following notation for the matrix elements of $\mathcal{V}$, $\mathcal{V}_{xy} \equiv \mathcal{V}_{(x)(y)} \equiv \bra{x}  \mathcal{V} \ket{y}$
% \begin{align}
%   \mathcal{V}_{xy} \equiv \mathcal{V}_{(x)(y)} \equiv \bra{x}  \mathcal{V} \ket{y} , 
% \end{align} 
we have  
\begin{widetext} 
\begin{align} 
    \mathbb{E}^{(0)}_{4} & = \sum_{\substack{p,q,r \\ q \not= p,r}} \frac{  \mathcal{V}_{0r} \mathcal{V}_{rq} \mathcal{V}_{qp}\mathcal{V}_{p0}}{ \Delta_{r} \Delta_{q} \Delta_{p}} - \sum_{p,q}  \frac{|\mathcal{V}_{p0}|^2 |\mathcal{V}_{q0}|^2 }{\Delta_{p}^2\Delta_{q}}  \nonumber  \\
   & = \sum_{j} \frac{ \mathcal{V}_{0(1_j)} \mathcal{V}_{(1_j)(2_{j})} \mathcal{V}_{(2_j)(1_{j})} \mathcal{V}_{(1_{j})0} }{ \Delta_{2_j} \Delta_{1_j}^2 } + \sum_{\substack{j,k \\ j\not= k}} \frac{ \mathcal{V}_{0(1_j)} \mathcal{V}_{(1_j)(1_j,1_k)} \mathcal{V}_{(1_j,1_k)(1_{j})} \mathcal{V}_{(1_{j})0} }{ \Delta_{1_j,1_k} \Delta_{1_j}^2 } \nonumber 
    \\
   & \;\;\; + \sum_{\substack{j,k \\ j\not= k}} \frac{  \mathcal{V}_{0(1_k)} \mathcal{V}_{(1_k)(1_j,1_k)} \mathcal{V}_{(1_j,1_k)(1_{j})}\mathcal{V}_{(1_{j})0}}{\Delta_{1_k} \Delta_{1_j,1_k} \Delta_{1_j}} + \sum_{\substack{j,k,l \\ k \not= j,l}} \frac{ \mathcal{V}_{0(1_l)}  \mathcal{V}_{(1_l)(1_k)} \mathcal{V}_{(1_k)(1_j)} \mathcal{V}_{(1_j)0} }{ \Delta_{1_l} \Delta_{1_k} \Delta_{1_j}} 
   - \sum_{j,k} \frac{  | \mathcal{V}_{(1_{j})0}|^2 | \mathcal{V}_{(1_{k})0}|^2}{ \Delta_{1_j}^2 \Delta_{1_k} } \nonumber \\ 
   & = N(N-1) \sum_{j} \left[  \frac{ V_{0j}^2 V_{j0}^2}{\Delta_{j}^3 } 
   + \sum_{k \not= j} \frac{  V_{0j} V_{0k} V_{k0} V_{j0}}{ \Delta_{j}^2 (\Delta_j + \Delta_k)  } +  \frac{  V_{0k} V_{0j} V_{k0} V_{j0}}{  \Delta_{j} \Delta_{k}  (\Delta_j + \Delta_k)} \right]  + N \sum_{\substack{j,k,l \\ k \not= j,l}} \frac{ V_{0l} V_{lk} V_{kj} V_{j0}}{\Delta_{l}\Delta_{k} \Delta_{j}   }  - N^2 \sum_{j,k} \frac{  | V_{j0}|^2 | V_{k0}|^2 }{ \Delta_{j}^2 \Delta_{k} }  
\end{align} 
\end{widetext}
If we apply the condition that $V_{k0}$ and $V_{j0}$ commute, which is true for both the RWA and product (bare-coupling) interactions, then we can rearrange this as 
\begin{align} 
    \mathbb{E}^{(0)}_{4} 
   & = N(N-1)\sum_{j,k} \left[ \frac{  V_{0j} V_{0k} V_{k0} V_{j0}}{ \Delta_{j}^2 \Delta_k  }  \right] \nonumber \\
   & \;\;\; + N\sum_{\substack{j,k,l \\ k \not= j,l}} \frac{ V_{0l} V_{lk} V_{kj} V_{j0}}{\Delta_{l}\Delta_{k} \Delta_{j}   }  - N^2 \sum_{j,k} \frac{  | V_{j0}|^2 | V_{k0}|^2 }{ \Delta_{j}^2 \Delta_{k} } \nonumber \\
   & = N \hat{E}^{(0)}_{4} 
   + N(N-1) \sum_{j,k}  V_{0j} \left( \frac{  [ V_{0k},  V_{j0}]  }{ \Delta_{j}^2 \Delta_k  } \right) V_{k0} . 
   \label{EE04}
\end{align} 
We see from this expression that if $[V_{0k},V_{j0}] = 0$ (in addition to $[V_{k0},V_{j0}] = 0$) then the second term vanishes and the nonlinearity generated by the ensemble is exactly $N$ times the nonlinearity generated by a single emitter: 
\begin{align} 
    \mathbb{E}^{(0)}_{4} 
   & = N \hat{E}^{(0)}_{4} , \;\;\;\; [ V_{0k},  V_{j0}] = 0  . 
\end{align} 
This commutator, $[ V_{0k},  V_{j0}]$, is zero for the bare emitter/field interaction but not for the RWA interaction. The procedure for calculating nonlinearities for an ensemble of emitters is summarized in Table~\ref{tab3}. 

\begin{table}[t]
\textbf{Procedure to calculate nonlinearities for an ensemble} \\
\vspace{1mm}
\hrule
    \begin{enumerate}
    \item Determine the emitter matrix elements of the interaction operators for a single emitter. These matrix elements are proportional to annihilation and creation operators for the field modes
    \item Determine the emitter dressed states 
    \item Transform the matrices for the interaction operators to this dressed-state basis 
    \item Use Eqs.(\ref{rNVj0})-(\ref{1l1j}) to calculate the matrix elements of the collective interaction(s) from those of the single-emitter interactions determined in step 3. 
    \item Substitute the elements of the collective interaction matrices into the formulae in Table~\ref{tab1} 
    %\\
    %4 & If desired, use Eq.(\ref{sigman}) to calculate the lindblad operators 
\end{enumerate}
\hrule 
\caption{The dressed states referred to here are the eigenstates of the emitter Hamiltonian including the classical driving fields. Note that the only difference in the above procedure to that for a single emitter is the addition of step 4. .\label{tab3}}
\end{table}

\subsection{Bound on the emitter/field coupling for the \\ ``bare-coupling" regime}
\label{barecoup}

As discussed above, well-defined nonlinearities can only be generated so long as the coupling between the emitter levels and the field(s) are in the perturbative regime, which places a bound on the size of this coupling, and in turn on the size of the nonlinearities. For an ensemble, as we have seen, the coupling between the ensemble ground and first symmetric excited states is $\sqrt{N}$ times the coupling between the ground state of a single emitter and each of its excited states. The crucial question that we must answer is whether the perturbative bound applies to the single emitter coupling or to the resulting ensemble coupling. The latter would place a much more restrictive bound on the emitter coupling and prevent the size of the nonlinearities scaling with the size of the ensemble. Here we answer this question for the situation of ``bare-coupling'' between the ensemble and the field(s) in which the perturbation parameter is the coupling rate divided by the transition frequency, the interaction operator for each field mode is Hermitian, and the Hamiltonian is given by  Eq.(\ref{HA}). 

As in Section~\ref{sec2H} we consider a single field mode coupled to the ensemble. Denoting the Hermitian operator of the mode that couples to the emitters by $\Lambda$, we examine the subspaces defined by each of the eigenvectors of $\Lambda$. To this end we can write $H_{\ms{B}}$ (Eq.(\ref{HA})) as 
\begin{align} 
    \mathcal{H}  = H_0 +  G \Lambda + H_{\ms{f}} 
\end{align} 
where $H_0$ and $G$ are operators of the emitters. We define the eigenvalues and eigenstates of the mode operator $\Lambda$ by $\Lambda \ket{\lambda} = \lambda \ket{\lambda}$. For each subspace defined by the eigenvector $|\lambda\rangle$ the Hamiltonian of the emitter is 
\begin{align} 
    \mathcal{H}_\lambda  = H_0 + \lambda G = \sum_j H_0^{(j)} + \lambda \sum_j G_j = \sum_j H_{\lambda}^{(j)}
    \label{HlamEns}
\end{align} 
where 
\begin{align}
    H_{\lambda}^{(j)} = H_0^{(j)} + \lambda G_j
    \label{Hlamj}
\end{align} 
is a Hamiltonian for emitter $j$, with $H_0^{(j)}$ and $G_j$ the Hamiltonian and interaction operator for that emitter. 

From Eqs.(\ref{HlamEns}) and (\ref{Hlamj}) it is clear that in each of the eigenspaces of the mode interaction operator, $\Lambda$, the perturbation problem breaks up into $N$ entirely separate perturbation problems, one for each emitter. If the eigenvectors of each emitter are denoted by $\ket{m_\lambda^{(j)}}$, then the tensor product of these eigenvectors, 
\begin{align}
    \ket{m_\lambda} \equiv \prod_j \hspace{-4.1mm}{\textstyle\otimes}\hspace{1mm}  \ket{m_\lambda^{(j)}} 
\end{align} 
is a perturbed eigenvector of the ensemble. Not all eigenstates of the ensemble are of the above form. However, the tensor product of the ground state for each emitter is the ground state of the ensemble. Further, the perturbation expansion for each emitter is valid under the condition (see Eq.(\ref{lambound}))
\begin{align}
    \lambda \ll \frac{\Delta}{\sqrt{\langle n \rangle}} 
    \label{lamllD}
\end{align} 
where $\Delta$ is the frequency scale of the emitter transitions, $\langle n \rangle$ is the mean number of photons in the field that couples at rate $\lambda$, and we are assuming that the matrix elements of $G$ are order unity. Consequently the perturbative expansion for the ground state of the ensemble \textit{is also valid under this condition}. Recall that if we perform the perturbative expansion in the symmetric subspace of the ensemble, as detailed in Sections~\ref{tlsym} and~\ref{Arbsym}, then the condition for validity is 
\begin{align}
    \lambda \ll \frac{\Delta }{\sqrt{N \langle n \rangle} } . 
    \label{lamllDrN}
\end{align} 
It is natural to assume that this latter relation is both necessary and sufficient for the perturbative expansion to be valid. But strictly speaking it is only a sufficient condition. It is possible, although unlikely, that the perturbative expansion happens to give the correct states even when the perturbative parameter is not small. Our analysis above shows that this is precisely what happens for the bare emitter/field interaction. For the ground state for each value of $\lambda$ the perturbation expansion is not confined to the regime of Eq.(\ref{lamllDrN}); it is valid whenever the condition in Eq.(\ref{lamllD}) is satisfied. As a result, nonlinearities generated by ensembles in the ``bare coupling" regime scale as $N$ and such ensembles can generate giant nonlinearities. 

\begin{figure}[t]
\centering
\leavevmode\includegraphics[width = 1 \columnwidth]{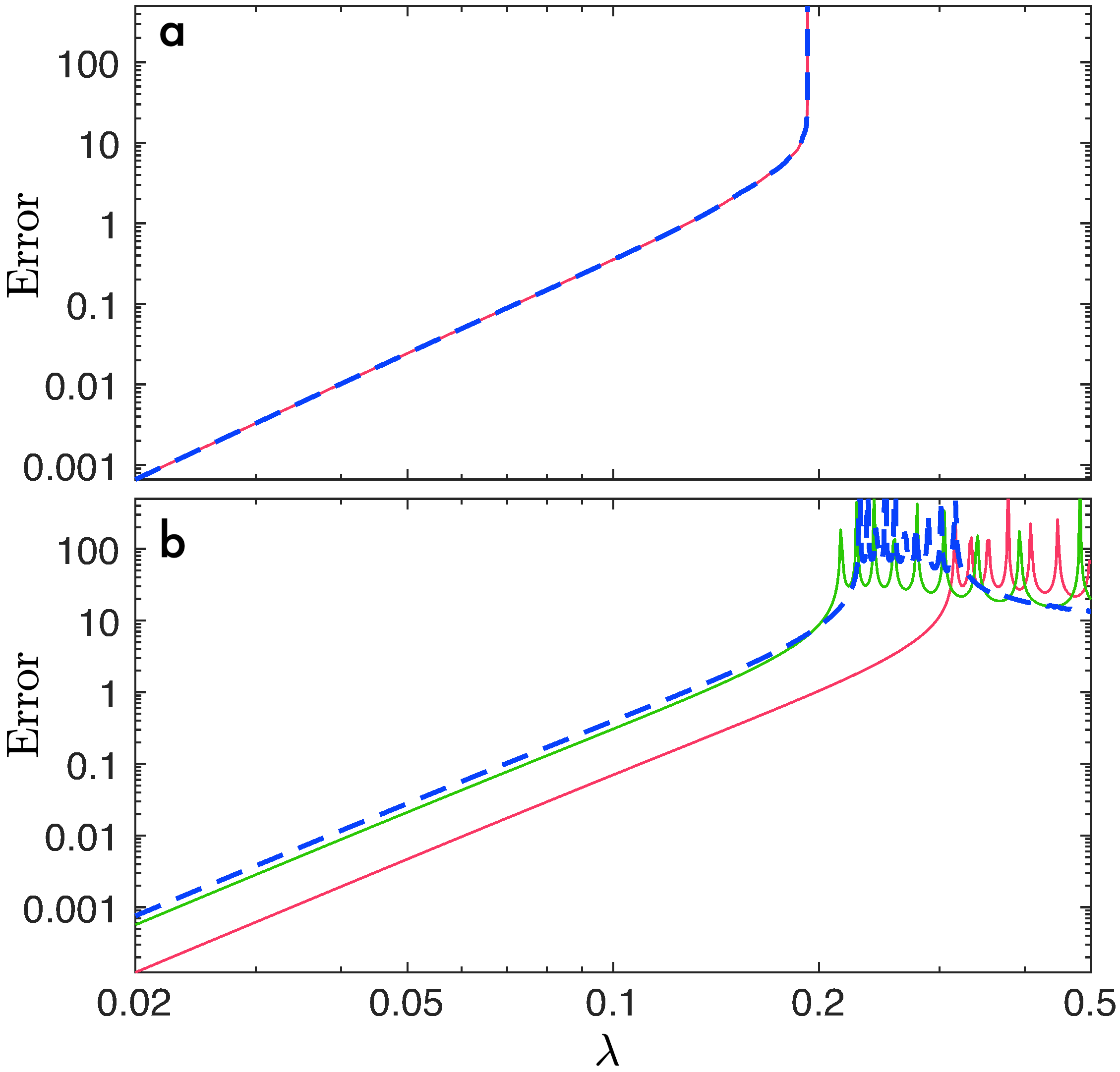}
\caption{(Color online) Numerical results that show the striking difference in behavior between nonlinearities generated by ensembles in the ``bare coupling" regime and those generated in the regime of the ``rotating-wave" approximation. Each curve is the sum of the squared error between the true eigenvalues in the ground state manifold and those given by truncating the TIPT expansion at 4$^{\msi{th}}$ order, for two level emitters. The red curve is for a single emitter ($N=1$), the blue curve is for 10 emitters ($N=10$). (a) Bare coupling: the blue and orange curves are identical. (b) Rotating-wave regime: the error increases with $N$ approximately as it does when $\lambda$ is increased by a factor of $N^{1/6}$, as expected (green curve).} 
\label{fig:Nscaling} 
\end{figure}

\subsection{Bound on the emitter/field coupling for the RWA regime}
\label{RWAbound}

In the section above we were able to show that for the ``bare-coupling" regime the perturbative expansion remained valid so long as it was valid for each emmitter independently, despite the fact that the coupling between the ensemble ground state and the first symmetric excited states was no longer perturbative. We now turn to the question of whether this is true for the RWA regime. Unfortunately the argument that we used for the bare-coupling regime does not hold for the RWA regime. Since we have not found a way to answer this question analytically we resort to a numerical calculation. 

We use as our example the simplest ensemble, that of $N$ two-level emitters that we analyzed for the case of the RWA regime in Section~\ref{tlsym}. For this ensemble we calculate the eigenvalues and eigenvectors for the ground-state subspace for both the RWA regime and the bare-coupling regime. We also calculate the eigenvalues of the effective Hamiltonians for the field that are generated in each of these regimes, but only to fourth order. As the coupling strength is increased, the difference between the $4^{\msi{th}}$-order Hamiltonian and the actual effective Hamiltonian that is generated, determined by the eigenvalues in the ground-state subspace, will increase. If it is the coupling between the ensemble ground-state and the symmetric states that determines the perturbative regime, then this error will increase with $N$, but if it is only the emitter/field coupling for single emitters that determines this regime, as we have shown is the case for bare-coupling, then this error will not change with $N$. We measure the error as the sum of the absolute value of the fractional difference between the eigenvalues of the exact effective Hamiltonian and the effective Hamiltonian calculated to fourth order.  

In Fig.~\ref{fig:Nscaling}a we plot the above-defined error as a function of the coupling rate, $\lambda$, for the bare coupling regime, and for two values of $N$, $N = 1$ and $N=10$. We see that both plots are identical, confirming our proof in Section~\ref{barecoup}. In Fig.~\ref{fig:Nscaling}b we plot the same error quantities for the RWA regime. We see that the RWA regime behaves quite differently to the bare coupling regime, with the error increasing when we go from $N=1$ to $N=10$. We can estimate this increase by examining the next term in the expansion for the effective Hamiltonian after $4^{\msi{th}}$ order. Since this term is $6^{\msi{th}}$ order in $\lambda$ and proportional to $N$, the increase in the error with $N$ should be approximately that resulting from increasing $\lambda$ by a factor of $N^{1/6}$. The error for the latter is plotted as the green curve in Fig.~\ref{fig:Nscaling}. We can conclude that for the RWA regime the bound on $\lambda$ is determined by the collective coupling, and is given by Eq.(\ref{lamllDrN}).  

Finally, it is important to note that even when the perturbation expansion is governed by the collective interaction, not all the coupling rates between the field and the emitter transitions have a bound that depends on $N$. It is only the coupling rates for transitions that involve the ground state that are restricted in this way. This means, for example, that in the Schmidt-Imamo{\=g}lu scheme, the coupling rate $\lambda$ is bounded as (see Eq.(\ref{lambound}))
\begin{align}
   \lambda \ll \frac{\Omega}{\sqrt{N \langle a^\dagger a \rangle}} 
   \label{lambound4l}
\end{align} 
but $\mu$ is bounded only as 
\begin{align}
   \mu \ll \frac{\Delta}{\sqrt{\langle b^\dagger b \rangle}}
   \label{mubound4l}
\end{align}
This fact impacts the scaling of the effective nonlinearity, as will be seen in the next section. 

\subsection{Scaling of the Kerr nonlinearity in the RWA regime: two-level systems vs. the Schmidt-Imamo{\=g}lu 4-level scheme}
\label{KerrScale}

We have seen that in the RWA regime, the bounds on all coupling rates for transitions that involve the ground state scale as $1/\sqrt{N}$, where $N$ is the size of the ensemble. The bounds on the coupling rates for all other transitions, however, are essentially unaffected by the size of the ensemble. We can now substitute the bounds for the coupling rates into the expressions for the sizes of the induced nonlinearities to see how the resulting bounds on the nonlinearities scale with the number of emitters. For the ensemble of two-level emitters the bound on the self-kerr nonlinearity is  
\begin{align}
   \kappa_{\ms{s}} = \frac{ N \lambda^4 }{ \Delta^3 } 
  & \ll \frac{ \Delta }{N \langle a^{\dagger} a \rangle^2} ,  
\end{align}
where we have used Eqs.\ (\ref{sKerr}) and (\ref{limrN}). 
% \begin{align}
%   \kappa_{\ms{s}} =  \frac{  \lambda^4  \mathbb{E}^{0}_4 }{\hbar}
%   & = \frac{ N \lambda^4 }{ \Delta^3 } (a^{\dagger} a)^2 +  \frac{ N(N-1) \lambda^4 }{ \Delta^3 }  a^{\dagger} a  \nonumber \\
%   & \ll \frac{ \Delta }{N} \frac{(a^{\dagger} a)^2}{\langle a^{\dagger} a \rangle^2} +  \frac{ \Delta (N-1)}{N} \frac{a^{\dagger} a}{\langle a^{\dagger} a \rangle^2}
% \end{align}
For an ensemble of 4-level emitters employing the Schmidt-Imamo{\=g}lu scheme, we first calculate the term that gives the cross-Kerr nonlinear Hamiltonian for an ensemble, being 
\begin{align}
 \lambda^2 \mu^2 \mathbb{E}^{(0)}_{22} & = \sum_{\substack{p,q,r \\ q \not= p,r}} \frac{  \mathcal{V}_{0r} \mathcal{X}_{rq} \mathcal{X}_{qp}\mathcal{V}_{p0}}{ \Delta_{r} \Delta_{q} \Delta_{p}}  \nonumber \\ 
    & = \sum_{\substack{j=1,2 \\ k=1,2}} \frac{ \mathcal{V}_{0(1_k)} \mathcal{X}_{(1_k)(1_{3})} \mathcal{X}_{(1_3)(1_{j})} \mathcal{V}_{(1_{j})0} }{ \Delta_{1_3} \Delta_{1_k}\Delta_{1_j} }  \\
    & = -\frac{N \hbar \lambda^2 \mu^2 }{\Delta\Omega^2} a^\dagger a b^\dagger b  . 
    \label{KerrEns}
\end{align}
The bound on the rate of the cross-Kerr nonlinearity is then 
\begin{align}
  \kappa & = \frac{N \lambda^2 \mu^2 }{\Delta\Omega^2}  \ll   \frac{ \Delta }{\langle a^\dagger a \rangle \langle b^\dagger b \rangle} , 
\end{align}
where we have used Eqs.\ (\ref{lambound4l}) and (\ref{mubound4l}). We see that the scaling of the maximal size of an effective Kerr nonlinearity depends on the nature of the ensemble. For the Schmidt-Imamo{\=g}lu scheme the bound on the kross-Kerr nonlinearity is independent of $N$, meaning that ensembles can generate the maximal strength of nonlinearity that can be produced by a single emitter. The maximal self-Kerr nonlinearity that can be generated by a ensemble of two-level systems, however, decreases with the size of the ensemble. Thus two-level systems \textit{do} become more linear as their number increases so long as we are in the deep quantum regime. Driven multi-level ensembles need not, as the Schmidt-Imamo{\=g}lu scheme shows. 

\subsection{The RWA regime: emergence of giant nonlinearities for semiclassical fields} 
\label{semiclass}

We saw in Section~\ref{barecoup} that in the bare coupling regime ensembles can generate nonlinearities that scale linearly with the number of emitters. The reason for this was that the operator by which each field mode couples to the system is Hermitian. This is not the case in general for the RWA regime. We now consider what happens when a field mode is in a coherent state with large photon number, and in particular the action of the creation operator on that state. Defining $\ket{\alpha}$ as a coherent state with amplitude $\alpha$, we have 
%\begin{align}
%    \bra{\alpha} (a^\dagger -  \alpha^*) \ket{\alpha} = 0 , 
%\end{align}
%and 
\begin{align}
    \bra{\alpha} (a - \alpha) (a^\dagger -  \alpha^*) \ket{\alpha} = 1 . 
\end{align}
This implies that  
\begin{align}
   a^\dagger  \ket{\alpha} =  \alpha^* \ket{\alpha} + \ket{\alpha}_\perp = \alpha^* \left(  \ket{\alpha} + \frac{\ket{\alpha}_\perp}{\alpha^*} \right) . 
\end{align}
where the state $ \ket{\alpha}_\perp$ is normalized. (It is also orthogonal to $\ket{\alpha}$.) Thus  
\begin{align}
    \lim_{|\alpha| \rightarrow \infty}  a^\dagger  \ket{\alpha}  = \alpha^*\ket{\alpha} . 
\end{align}
The RWA interaction between a field mode with annihilation operator $a$ and the emitter transition $j\leftrightarrow k$ is   
\begin{align}
   H_{jk} = \hbar \lambda \left(  a \sigma^\dagger + a^\dagger \sigma \right)  . 
\end{align}
For large $\alpha$ the action of this interaction on $\ket{\alpha}$ is 
\begin{align}
   H_{jk} \ket{\alpha} \approx \hbar \lambda |\alpha| \left(  \sigma^\dagger e^{-i\theta} +  \sigma e^{i\theta}  \right) \ket{\alpha} . 
\end{align} 
where $\theta$ is the phase of $\alpha$. This interaction thus acts like it is the product of a Hermitian operator for the mode (whose eigenstates are $\ket{\alpha}$ for $|\alpha| \gg 1$) and the Hermitian operator $\sigma_{\theta} =  \sigma^\dagger e^{-i\theta} +  \sigma e^{i\theta}$ for the emitter. The interaction Hamiltonian for the ensemble now has the same structure as that of the bare coupling regime, and by the analysis in Section~\ref{barecoup} the nonlinearities generated by ensembles in the RWA regime, and thus EIT, scale with the size of the ensemble in the semiclassical regime.  

\section{Calculating nonlinearities for traveling-wave fields}
\label{travwave}

So far we have written the nonlinearities generated by emitters in terms of the coupling rates between the emitters and field modes, assumed to be discrete modes of a cavity. But we often wish to consider the generation of nonlinearities for traveling-wave fields such as laser beams or pulses. For this we need to be able to calculate the Rabi frequency from the laser power (for a continuous beam) or the number of photons in a laser pulse. Recall that the Rabi frequency is required to determine whether the emitter/field system is in the perturbative regime. 

The coupling energy (matrix element) between two emitter energy levels $j$ and $k$ is given by $\mu_{jk} E$ where $\mu_{jk}$ is the dipole matrix element between the two levels and $E$ is the electric field amplitude at the location of the emitter. The Rabi frequency (the coupling rate) for the transition is this coupling energy divided by $\hbar$: 
\begin{align}
  \Omega_{jk} = \frac{\mu_{jk} E}{\hbar} . 
\end{align}
The intensity of an electromagnetic plane wave with amplitude $E$ is $I = \varepsilon_0 c |E|^2/2$. The Rabi frequency induced by a laser beam with power $P$ and cross sectional area $A$ is therefore 
\begin{align}
  \Omega_{jk} = \frac{\mu_{jk}}{\hbar} \sqrt{\frac{2P}{\varepsilon_0 c A}}. 
\end{align}
For a narrow-band pulse of length $L$ containing $n_{\ms{p}}$ photons the average power is $P = \hbar \omega n_{\ms{p}} c/L$ and so the average Rabi frequency is 
\begin{align}
  \Omega_{jk} = \frac{\mu_{jk}}{\hbar} \sqrt{\frac{2\hbar \omega n_{\ms{p}} }{\varepsilon_0 L A}} = \mu_{jk} \sqrt{\frac{2 \omega n_{\ms{p}} }{\varepsilon_0 \hbar V}}  
  \label{Rabifreq}
\end{align}
where $V = LA$ is the pulse volume. 

The single-emitter coupling rate, $g$, is then 
\begin{align}
    g_{jk} = \frac{\Omega_{jk}}{\sqrt{n_{\ms{p}}}} =  \mu_{jk} \sqrt{\frac{2 \omega }{\varepsilon_0 \hbar V}}
\end{align}

\subsection*{Nonlinear susceptibility}

So far, we have characterized the nonlinearities generated by ensembles as terms in a Hamiltonian for one or more cavity modes. For situations in which the fields are traveling waves in free space, the nonlinearities are instead often quoted as \textit{nonlinear susceptibilities} and are written in terms of the number density of emitters. It is simple to translate between the rate constant that multiplies a nonlinear term in a Hamiltonian for cavity modes and the resulting nonlinear susceptibility. To do so you first replace each of the emitter/field interaction rates that we use in our analysis ($\lambda, \nu, \mbox{etc.}$) with the absolute value of the dipole moment of the corresponding emitter transition, divided by $\hbar$. Thus if mode $a$ interacts with transition $j \leftrightarrow k$ with rate $\lambda$, then you replace  
\begin{align}
    \lambda \rightarrow \frac{|\mu_{jk}| }{\hbar} ,
\end{align}
where $\mu_{jk}$ is the dipole moment for the transition $j \leftrightarrow k$. You then multiply the whole expression for the nonlinear rate constant by $\hbar/(\varepsilon_0 V)$ where $V$ is the mode volume. In fact, as we have shown above, the resulting nonlinear susceptibility will only be exactly proportional to $n_{\ms{d}}$, the number of emitters per unit volume, for nonlinearities generated using the ``bare-coupling''  regime or in the semi-classical limit for the RWA regime, because it is only in those cases that the nonlinearities are exactly proportional to the number of emitters. In other cases, there are corrections at higher orders in the perturbation parameter(s).  

As an example, converting the rate constant for the cross-Kerr nonlinear Hamiltonian generated by the Schmidt-Imamo{\=g}lu scheme, given in Eq. (\ref{KerrEns}), the nonlinear susceptibility generated by this scheme is  
\begin{align}
    \chi^{(3)} = \frac{n_{\ms{d}} |\mu_{12}|^2 |\mu_{34}|^2 }{\varepsilon_0 \hbar^3 \Delta\Omega^2}  ,  
\end{align}
in which the density of emitters, $n_{\ms{d}} = N/V$. 

Nonlinear susceptibilities are defined in terms of the equation of motion for the electric field. This equation is~\cite{Boyd08}
\begin{align}
    \nabla \times \nabla \times \mbox{\textbf{E}} + \frac{1}{c^2}\frac{\partial^2}{\partial t^2}\mbox{\textbf{E}}
= -\frac{1}{\varepsilon_0 c^2}\frac{\partial^2}{\partial t^2}\mbox{\textbf{P}} 
   \label{Eeq}
\end{align}
where the (nonlinear) polarization is 
\begin{align}
    \mbox{\textbf{P}} = \varepsilon_0 \sum_m \chi^{(m)}  \mbox{\textbf{E}}^m .
    \label{defChi3}
\end{align}
Here the constant $\chi^{(m)}$ is the $m^{\msi{th}}$-order nonlinear susceptibility. To determine $\chi^{(m)}$ from the nonlinear Hamiltonian one simply uses the Hamiltonian to derive Eq.(\ref{Eeq}). 

From Eqs.(\ref{Eeq}) and (\ref{defChi3}) a material whose only  susceptibility is $\chi^{(3)}$ has a refractive index of 
\begin{align}
    n_{\ms{r}} = \sqrt{1 + \frac{\chi^{(3)} |E|^2}{\varepsilon_0} } . 
    \label{indexnr}
\end{align}
For a self-Kerr nonlinearity $E$ is the amplitude of the field experiencing the refractive index, whereas for a cross-Kerr nonlinearity $E$ is the amplitude of a travelling wave with a different direction and/or frequency than that experiencing the refractive index. 

% \begin{align}
%     \varepsilon_0 E + \chi^{(1)} E + \chi^{(3)} E |E|^2 = \varepsilon_0 \left( 1 + \frac{\chi^{(1)}}{\varepsilon_0} + \frac{\chi^{(3)} |E|^2}{\varepsilon_0} \right)  E  \\
%     n_{\ms{r}} = \sqrt{1 + \frac{\chi^{(1)}}{\varepsilon_0} + \frac{\chi^{(3)} |E|^2}{\varepsilon_0} }
% \end{align}

\section{Potential of ensembles to generate giant nonlinearities}
\label{potstr}

We have seen that for ensembles in the regime of the rotating-wave approximation, and outside the semi-classical regime, the limitation on the coupling rate, $\lambda$, between the ground state and any excited states limits the strength of any nonlinearities to those that can be generated by a single emitter with the maximum allowed value for $\lambda$. Since the expressions for the size of the nonlinearities for a single emitter are always equal to one of the coupling rates multiplied by one or more perturbation parameters, these nonlinearities can never be larger than the coupling rate. One can only obtain strong coupling rates by using tightly confined cavities, and even then the nonlinear rate will be at least an order of magnitude smaller than the coupling rate. The only way to break this limit on the size of the  nonlinearities is to take advantage of the scaling with the ensemble size, $N$, which requires either working in the semi-classical regime or  the bare coupling regime. 

Note that we have not determined 
%\textit{where} the boundary is between the fully quantum regime in which nonlinearities are severely restricted and the semi-classical regime in which they scale freely with $N$. 
the amplitude required by a field or field to place the system in the semi-classical regime for the purposes of generating effective nonlinearities. Quite different behavior will be accessible for very weak fields if only 5 photons are required as opposed to $10^5$ or $10^{15}$. We will not explore this question further here but it is certainly an important topic for future work. 

We now compare our theoretical results to an experiment by Venkataraman, Saha, and Gaeta (VSG) in which they realize a scheme for generating a cross-Kerr nonlinearity using a three-level  emitter~\cite{Venkataraman13}. They obtain especially strong coupling between an ensemble of rubidium atoms and two narrow band laser beams (one a pulse and the other continuous) by confining a rubidium vapor inside a hollow optical fiber. The beams propagate down the hollow core of the fiber which has a diameter of $6~\mu\mbox{m}$. The length of the pulse is $L = 1.5~\mbox{m}$. The density of rubidium atoms is $n_{\ms{d}} = 2\times 10^{19}~\mbox{m}^{-3}$ so that $N = 8.5\times 10^{8}$ atoms interact with the pulse. The pulse, which couples the ground state to the first excited state, contains 20 photons, giving it an average power of $1\, \mbox{nW}$. The power of the continuous beam is $P = 10\, \mu \mbox{W}$. The phase shift imparted to the continuous beam by the pulse via the cross-Kerr nonlinearity can be determined from the refractive index which is calculated from the cross-Kerr nonlinear susceptibility, $\chi^{(3)}$, using Eq.(\ref{indexnr}). In~\cite{Venkataraman13} the authors find good agreement between their measured phase shift and the cross-Kerr susceptibility calculated using the standard semi-classical analysis. Calculating the Rabi frequency induced by the 20-photon pulse for a single emitter (Eq.(\ref{Rabifreq})) we obtain $\Omega = 8.9~\mbox{MHz}$. The Rabi frequency generated by the collective coupling, on the other hand, is $\sqrt{N}\Omega = 260~\mbox{GHz}$. Since the detuning between the transitions and the fields is $700~\mbox{MHz}$, the single-emitter coupling is in the perturbative regime, but the collective coupling is not. Thus the agreement between the experimental results and the semi-classical theory, achieving a nonlinearity enhanced by a factor of $N$, implies that as few as 20 photons is  sufficient, at least on a timescale of $L/c = 5~\mbox{ns}$, for the system to reach the semi-classical limit.  

% The strength (rate) of the Kerr nonlinearity for this scheme is 
% \begin{align} 
%     \kappa = \frac{N\lambda^2 \nu^2}{\Delta_1 \Delta_2}
% \end{align}

%\vspace{4mm}

\section{Conclusion}
\label{conc}

Optical nonlinearities are essential for many applications, including quantum information processing using photonic qubits. Understanding the scaling of nonlinearities with the number of emitters, $N$, is essential for constructing schemes that harness these nonlinearities. To this end, we introduced a method for calculating the optical nonlinearities generated by driven multi-level emitters and ensembles of independent emitters. We have shown that the scaling of these nonlinearities with $N$ is remarkably subtle: different regimes and different level structures lead to different scaling behavior, which in turn determines the size of the nonlinearities that can be generated.

In particular, in the rotating-wave approximation (RWA) regime, the size of the nonlinearities scales with $N$ when the fields are in the semi-classical regime, but is limited for single-photon fields due to the effect of collective coupling. This result implies that the ability of ensembles to generate "giant" nonlinearities for very weak fields will depend on the field strength at which the field/ensemble system makes the transition to the semi-classical regime. This question is an interesting topic for future work.

Outside the RWA regime, ensembles can generate nonlinearities for single-photon fields that scale as $N$. We suggest that this prediction be tested using, for example, Rydberg atoms~\cite{Graham19}, solid-state cavity-QED systems~\cite{Waks17, Fahey23, Zhong19} or superconducting circuits~\cite{Blais20}. 

The theory that we have developed here should allow a much more detailed understanding than previously possible of the size of the nonlinearities that can be generated by small and large numbers of emitters. This understanding can be expected to facilitate the manipulation of nonlinearities for quantum and classical technologies in a wide range of settings.

\appendix 

\section{Operators defined by Inner products of subsystem ``states"} 
\label{appA}

When considering a system consistng of two subsystems it is common to employ the state of one of the systems as a projector onto a subspace. Specifically, let us denote a set of basis states for subsystem $A$ by $\ket{k}_A$, $k = 0,1,\ldots, K$, and a set for subsystem B by $\ket{l}_B$, $l = 0,1,\ldots,L$. If subsystem A is in state $\ket{n}_A$ this defines a subspace consisting of all states of the form $\ket{n}_A \otimes \ket{l}_B$ for $l = 0,1,\ldots,L$. The projector onto this subspace is 
\begin{align}
    P_n =  \ket{n}_A \bra{n}_A \otimes I_B 
\end{align}
where $I_B$ is the identity operator for system B.
Consider an operator $C$ of the joint system, 
\begin{align}
    C = \sum_{kk'} \sum_{ll'} c_{kl,k'l'} \ket{k}_A \bra{k'}_A \otimes \ket{l}_B \bra{l'}_B
\end{align}
If we apply the projector $P_n$ to $C$ we get  
\begin{align}
    P_n C P_n = \ket{n}_A \bra{n}_A \otimes  \bra{n}_A C \ket{n}_A 
    \label{pncpn}
\end{align}
where 
\begin{align}
    \bra{n}_A C \ket{n}_A =  \sum_{ll'} c_{nl,nl'} \ket{l}_B \bra{l'}_B
\end{align}
is an operator that acts only in the space of B. While in terms of matrix operations the expression $\bra{n}_A C \ket{n}_A$ is not well-defined its meaning is clear in selecting out a sub-matix of $C$. If we are only interested in the action of that subblock on B (rather than the state of A), then we can discard the projector $\ket{n}_A \bra{n}_A$ on the right hand side of Eq.(\ref{pncpn}) above to leave us with $\bra{n}_A C \ket{n}_A$. This is what we do to arive at Eq.(\ref{sigman}). Note that the state $\ket{0}$ in Eq.(\ref{sigman}) represents a subspace although it is not purely a state of the emitter. It is given by the perturbation series in terms of the states of the emitter and operators on the mode(s). Using the perturbation series for $\ket{0}$ we can write the projector $\ket{0}\bra{0}$ explicitly in terms of emitter-state projectors and field operators. 

\section{Some elements of the TIPT eigenvector expansion} 
\label{AppEigvec}

Up to third order for two fields we have
\begin{align}
 \langle l_0 | n_{1} \rangle  & = \frac{V_{ln}}{ \Delta_{nl}}    , \;\;\; l \not= n \\
 \langle l_0 | n_{2} \rangle  &  =  \sum_{q\not=l,n}  \frac{V_{lq}V_{qn}}{ \Delta_{nl}\Delta_{nq}}    , \;\;\; l \not= n  \\
 \langle l_0 | n_{3} \rangle  &  =  \sum_{q\not=l,n}   \sum_{k\not=q,n}  \frac{V_{lq} V_{qk}V_{kn}}{ \Delta_{nl}\Delta_{nq}\Delta_{nk}} - \frac{1}{2}\sum_{q\not=n} \frac{V_{ln}V_{nq} V_{qn}}{ \Delta_{nl}\Delta_{nq}^2} \nonumber \\ 
 & \;\;\; -  \sum_{q\not= n}  \frac{ V_{ln} V_{nq} V_{qn}  }{\Delta_{nq} \Delta_{nl}^2 }   , \;\;\; l \not= n \\ 
 \langle l_0 | n_{11} \rangle  & = \sum_{q\not=l,n}  \frac{V_{lq} X_{qn} + X_{lq} V_{qn}}{ \Delta_{nl} \Delta_{nq}}  , \;\;\; l \not= n 
\end{align}
\begin{align} 
 \langle l_0 | n_{21} \rangle & =  \sum_{q\not=l,n} \sum_{p\not=q,n}    \frac{ \mathsf{ALLP}\mbox{\textbf{[}}\!\mbox{[} \, V_{lq}V_{qp} X_{pn}   \mbox{]}\!\mbox{\textbf{]}}  }{ \Delta_{nl} \Delta_{nq} \Delta_{np}} \nonumber \\
 & \:\:\:  - \sum_{q\not=n} \frac{\mathsf{ALLP}\mbox{\textbf{[}}\!\mbox{[} \, V_{ln} V_{nq} X_{qn}  \mbox{]}\!\mbox{\textbf{]}}}{ \Delta_{nl} \Delta_{nq}}   \left( \frac{1}{\Delta_{nl}} + \frac{1}{ 2\Delta_{nq}} \right)    , \;\;\; l \not= n 
\end{align} 
and for the special case of $l=n$ these are 
\begin{align}
   \langle n_0 | n_{2} \rangle & = - \frac{1}{2} \sum_{l\not= n} \frac{V_{nl}V_{ln}}{ \Delta_{nl}^2}    \\ 
    \langle n_0 | n_3 \rangle & = - \frac{1}{2}  \sum_{k,q\not=n}   \frac{ V_{nk}V_{kq}V_{qn} }{ \Delta_{nk}\Delta_{nq}} \left( \frac{1}{\Delta_{nk}} + \frac{1}{ \Delta_{nq}}  \right)   \\
    \langle n_0 | n_{11} \rangle & = - \frac{1}{2}  \sum_{k\not= n} \frac{ V_{nk}X_{kn} + X_{nk}V_{kn}  }{\Delta_{nk}^2} \\
    \langle n_0 | n_{21} \rangle & = - \frac{1}{2}\sum_{l,q\not=n}^N   
         \frac{ \mathsf{ALLP}\mbox{\textbf{[}}\!\mbox{[} \, V_{nl} V_{lq} X_{qn}  \mbox{]}\!\mbox{\textbf{]}} }{\Delta_{nl} \Delta_{nq}}\left( \frac{1}{\Delta_{nl}} + \frac{1}{ \Delta_{nq}}  \right)  
\end{align}

\end{document}

% \title{Supplement to ``Unified quantum theory of single-photon nonlinearities in ensembles of independent emitter''}

% \author{Kurt Jacobs}
% \affiliation{United States Army Research Laboratory, Adelphi, Maryland 20783, USA}
% \affiliation{Department of Physics, University of Massachusetts at Boston, Boston, Massachusetts 02125, USA}

% \author{Stefan Krastanov} 
% \affiliation{Department of Electrical Engineering and Computer Science, Massachusetts Institute of Technology, Cambridge, MA 02139, USA}

% \author{Mikkel Heuck}
% \affiliation{Department of Electrical and Photonics Engineering, Technical University of Denmark, 2800 Lyngby, Denmark}

% \author{Dirk R. Englund}% 
% \affiliation{Department of Electrical Engineering and Computer Science, Massachusetts Institute of Technology, Cambridge, MA 02139, USA}

% %\date{\today}

% \begin{abstract} 
% \end{abstract}

% \maketitle

\begin{center} \textbf{ Supplement to ``Unified quantum theory of single-photon nonlinearities in ensembles of independent emitter''} \\
\vspace{3mm} 
Kurt Jacobs, Stefan Krastanov, Mikkel Heuck, Dirk R. Englund 
\end{center}

Here we use our TIPT method to calculate the nonlinearities induced for a single cavity mode by a single two-level emitter in the rotating-wave regime. This is the situation in which an off-resonant two-level emitter induces a frequency shift for a cavity mode, often referred to as a \textit{dispersive} interaction between the emitter and the cavity. The ``dispersive interaction'' referred to is merely the first non-zero term in the TIPT expansion for the effective Hamiltonian for the field mode. 

The Hamiltonian for this system in the interaction picture is  
\begin{align}
    H = \hbar \left[ \frac{\Delta}{2} \sigma_z + \lambda (a \sigma^\dagger + a^\dagger \sigma) \right] 
\end{align}
in which the operators have their usual meanings. Using the matrix representation of the tensor product we can write the interaction operator $V$ as 
\begin{align}
    V = \hbar \left( \begin{array}{cc}
       0  & a \\
        a^\dagger  & 0
    \end{array}\right) 
\end{align}
Labelling the ground state of the emitter as 0 and the excited state as 1, we have 
\begin{align} 
    V_{00} & = V_{11} = 0 , \\ 
    V_{01} & = \hbar a^\dagger = V_{10}^\dagger   
\end{align}
and we note that $\Delta_1 = E^{(0)}_0 - E^{(1)}_0 = - \Delta$. Note also that because of the way the Pauli matrices are defined, the first index of the matrix elements of $V$ gives the collumn and the second index the row. The terms in the eigenvalue expansion for the perturbed ground state are  
\begin{align}
    \hat{E}_{2}^{(0)} & = \sum_{l\not= 0} \frac{ V_{0l}V_{l0}  }{\Delta_l} = \frac{\hbar}{\Delta} a^\dagger a ,  \\
    \hat{E}_{3}^{(0)} & = -\sum_{k,l\not= 0} \frac{ V_{0k}V_{kl}V_{l0} }{\Delta_k \Delta_l} = 0 , \\
    \hat{E}_{4}^{(0)} & = \sum_{\substack{l,k,q \not= 0 \\ q \not= k,l}}  \frac{ V_{0k}V_{kq}V_{ql}V_{l0} }{\Delta_q \Delta_k \Delta_l} - \sum_{k,l\not= 0}  \frac{ V_{0k}V_{k0}V_{0l}V_{l0} }{ \Delta_k^2 \Delta_l}  = \frac{\hbar}{\Delta^3} a^\dagger a a^\dagger a . 
\end{align}
The effective Hamiltonian for the ground-state subspace is therefore 
\begin{align}
    H_{\ms{eff}} & = \hat{E}^{(0)} = \sum_j \lambda^j \hat{E}^{(0)}_j = \hbar \lambda \left[ - \left( \frac{\lambda}{\Delta} \right) a^\dagger a  +  \left( \frac{\lambda}{\Delta} \right)^3 (a^\dagger a)^2 + \ldots \right] 
\end{align}
The first term is the familiar dispersive frequency shift of the cavity mode, and the second term is a self-Kerr nonlinearity. It is clear from the pattern of the expansion coefficients that all terms with odd powers of the mode opperators vanish. The effective Hamiltonian thus has the form 
\begin{align}
    H_{\ms{eff}} & = \hbar \lambda \sum_k c_{2k} \left(\frac{\lambda}{\Delta}\right)^{2k-1} (a^\dagger a)^{k} 
\end{align}
Using the recursion relations a computer code can easily calculate the coefficents $c_j$ to high order. We find that 
\begin{align}
    c_2 & = -1 , \\
    c_4 & = 1, \\
    c_6 & = -2 , \\
    c_8 & = 5 , \\
    c_{10} & = -14 , \\
    c_{12} & = 42 . 
\end{align}